\documentclass[11pt]{article}

\usepackage{geometry}
\usepackage{amsfonts,amssymb,amsmath,amsthm}
\usepackage{epsfig}

\newtheorem{Model}{Error Model}

\newcommand{\ncd}{\newcommand}
\ncd{\QCcns}{$QC_{\cal{C}}$}
\ncd{\QCc}{$QC_{\cal{C}}\;$}
\begin{document}

\title{A fault-tolerant one-way quantum computer}

\author{R. $\mbox{Raussendorf}^1$, J. $\mbox{Harrington}^2$ and
  K. $\mbox{Goyal}^1$   
\vspace{1.5mm}\\
$\mbox{}^1${\em{\small{Institute for Quantum Information, California
      Institute of Technology,  
Pasadena, CA 91125}}} \\
$\mbox{}^2${\em{\small{Los Alamos National Laboratory, Biological and
      Quantum Physics,  
MS D454, Los Alamos, NM 87545}}} }

\maketitle

\begin{center}
\parbox{0.8\textwidth}{\small{\noindent We describe a fault-tolerant
    one-way quantum computer on cluster states in three
    dimensions. The presented scheme uses methods of topological error
    correction resulting from a link between cluster states and
    surface codes. The error threshold is 1.4\% for local
    depolarizing error and 0.11\% for each source in an error model
    with preparation-, 
    gate-, storage- and measurement errors.}}
\end{center}

\section{Introduction}

A quantum computer as a physical device has to cope with imperfections
of its hardware. Fortunately, it turns out that arbitrary large
quantum computations can 
be performed with arbitrary accuracy, provided the error level of the
elementary components of the quantum computer is below a certain
threshold. This is the content of the threshold theorem for quantum
computation \cite{TT1,TT2,TT3,TT4}. The threshold theorem also provides
lower bounds to 
the error threshold which are in the range between $10^{-10}$ and
$10^{-4}$, depending on the error
model. It thus appears that there is a gap  
between the required and the currently available accuracy of quantum
operations, and it invites narrowing from both 
the experimental and the theoretical side. In this context,
significant progress has been made in \cite{Kn2} where a threshold
estimate in the percent range has been demonstrated. For
experimentally viable quantum computation there is a further
desideratum besides a high error threshold.  With the
exception of certain schemes for topological quantum computation
\cite{Kit1, Pr, Mo}, the price 
for fault-tolerance 
is an overhead in quantum resources.  This overhead should be moderate.  
 
Here we describe a fault-tolerant version of the one-way quantum
computer (\QCcns). The \QCc is a scheme for universal quantum
computation by one-qubit measurements on cluster states
 \cite{QCc}. Cluster states  \cite{BR}  consist of qubits arranged on a two- or
three-dimensional lattice and may be created by a nearest-neighbor Ising 
interaction. Thus, for the \QCcns, only nearest-neighboring qubits need
interact and furthermore, only once at the beginning of the
computation. For this scenario in three  
dimensions we present methods of error correction and a threshold value.

The existence of an error threshold for the \QCc
has previously been established \cite{FTQCc, FTQCc2, FTQCc2b} and
threshold estimates have been obtained \cite{FTQCc2, FTQCc2b}, by
mapping to the circuit 
model. Here we take a different path. We make use of topological error 
correction capabilities that the cluster states naturally
provide \cite{LRE} and which can be linked to surface codes
\cite{Kit1,Kit2}. The main design tool upon which 
we base our construction are engineered lattice defects which
are topologically entangled.  

\begin{figure}
  \begin{center}
    \epsfig{file=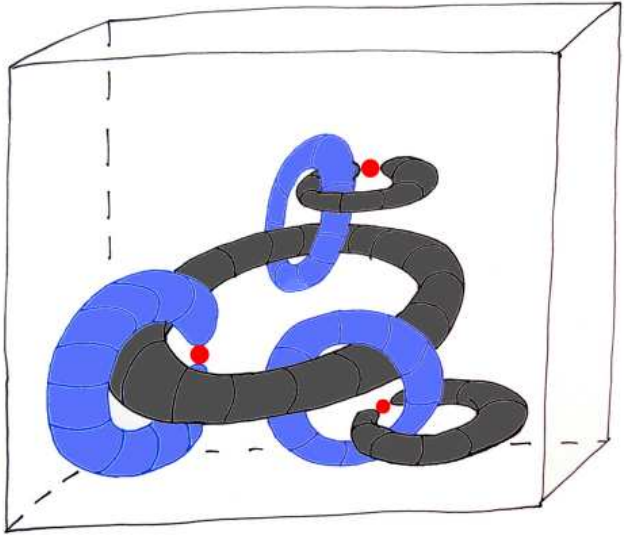, width=9cm}
    \caption{\label{scheme}Topological error correction for the
    \QCcns. The three-dimensional cluster state, shown as a block, is
    structured 
    by lattice defects which exist in two kinds: primal and
    dual. The defects  wind around another. Some of them
    hold singular qubits (red) which realize the non-Clifford part of
    the quantum computation.}
  \end{center}
\end{figure}

The picture is the following: quantum
computation is performed on a three-dimensional cluster state via a
temporal sequence of one-qubit measurements. The cluster lattice is
subdivided into three regions, $V$, $D$ and $S$. The set $S$ comprises
the `singular' qubits which are measured in an adaptive basis. The
quantum computation happens essentially there. The sets $V$ and $D$
are to distribute the correct quantum correlations among the qubits of
$S$. $V$ stands for `vacuum', the quantum correlations can propagate
and spread freely in $V$.  $D$ stands for `defect'. Quantum
correlations cannot penetrate the defect regions. They either end in
them or wrap around them. In both cases, the defects {\em{guide}} the
quantum correlations. $V$ and $D$ are distinguished by the bases in
which the respective cluster qubits are measured. The region $V$ fills
most of the cluster. Embedded in $V$ are the defects ($D$) most of which
take the shape of loops. These loops are topologically entangled with
another. Further, there are defects in the shape of ear clips which
each hold an $S$-qubit in their opening. Such a defect and the belonging
$S$-qubit form again a loop. These are the only
locations where the $S$-qubits occur. See Fig.~\ref{scheme}. \medskip

There are two codes upon which we base our construction, the planar code
\cite{Kit2} and the concatenated $[15,3,1]$ quantum Reed-Muller code
\cite{Kn,magic}. 
Both codes individually have their strengths and weaknesses, but they
can be advantageously combined. The planar code has a
relatively high error threshold of about 11\% \cite{RPGM}, and the
symmetries of 
its stabilizer fit well with cluster states. Moreover, fault-tolerant
data storage with the planar code and the creation of long range
entanglement among planar code qubits are easily accomplished in
three-dimensional cluster states, via a bcc-symmetric pattern of
one-qubit measurements. Therefore, it is suggestive to base
fault-tolerant cluster state computation on this code. However, the
planar code is not well suited to non-Clifford operations which are 
essential for universal quantum computation.   

Now, in cluster- and graph state quantum computation, the non-Clifford
part of the circuit is implemented by destructive measurements of the 
observables $(X\pm Y)/\sqrt{2}$ (The Clifford part is implemented by 
$X$-, $Y$- and $Z$-measurements.). The Reed-Muller quantum code is
well suited for fault-tolerant
quantum computation via local measurements because the measurements of
the encoded observables $(\overline{X}\pm \overline{Y})/\sqrt{2}$
(and of $\overline{X}$,  $\overline{Y}$, $\overline{Z}$ besides) are
accomplished fault-tolerantly by the respective measurements on the bare
level---and bare level measurements is what we are allowed to do in
the \QCcns. If we could assume we had 
given a graph state state as algorithmic resource where
each cluster qubit was encoded with the concatenated Reed-Muller code  
and where noise acted locally on the bare level, then fault-tolerant
quantum computation were trivial to achieve. However, is not obvious how to
create such an encoded graph state affected by local noise only. But
this is just what the topological error correction in cluster states
can do.

This paper is organized as follows. In
Section~\ref{objects}, we introduce the ingredients required for the
error correction mechanisms we use, namely cluster states, the 
planar code and the 15-qubit Reed-Muller quantum code. In
Section~\ref{MeaP} the 
measurement pattern used for fault-tolerant cluster state 
quantum computation is described, and in Sections~\ref{homology} -
\ref{Elm} it is explained. Specifically, in Section~\ref{homology} we
describe the physical objects  
relevant for the discussed scheme---defects, cluster state
quantum correlations, errors and syndrome bits---in the language of
homology. In Section~\ref{Elm} we introduce the techniques for
structuring quantum correlations via topological entanglement of
lattice defects. Our error models are stated in Section~\ref{EM} and
the fault-tolerance threshold is derived in 
Section~\ref{MECTV}. The overhead is estimated in Section~\ref{O}. We
discuss our results in Section~\ref{Disc}. 

\section{Cluster states and quantum codes}
\label{objects}

This section is a brief review of the ingredients for the described
fault-tolerant \QCcns. 

\paragraph{Cluster states.} A cluster state is a stabilizer state of
qubits, where each qubit occupies a site
on a $d$-dimensional lattice ${\cal{C}}$. Each site $a \in {\cal{C}}$
has a neighborhood $N(a)$ which consists of the lattice sites with the
closest spatial distance to $a$. Then, the cluster state
$|\phi\rangle_{\cal{C}}$ is---up to a
global phase---uniquely defined via the generators $K_a$ of its
stabilizer
\begin{equation}
  \label{corr}
  K_a:=X_a \bigotimes_{b \in N(a)}Z_b,\;\; \forall a \in {\cal{C}}, 
\end{equation} 
i.e., $|\phi\rangle_{\cal{C}} = K_a |\phi\rangle_{\cal{C}}$.
Here, $X_a$ and $Z_b$ are a shorthand for the Pauli operators
$\sigma_x^{(a)}$ and $\sigma_z^{(b)}$ that we use throughout the
paper. We refer to the generators $K_a$ of the cluster state
stabilizer as the elementary cluster state quantum correlations.

In this paper, we will use as the lattice  underlying the cluster
state a bcc-symmetric lattice in tree dimensions. That is, the
location of cluster qubits is given by lattice vectors
\begin{equation}
  \label{e/o}
        \begin{array}{lcr}
        \{(o\mbox{[dd]},e\mbox{[ven]},e), (e,o,e), (e,e,o)\},&& 
	\mbox{odd qubits},\\
        \{(e,o,o),(o,e,o),(o,o,e)\}, && \mbox{even qubits}. 
        \end{array}     
\end{equation}
We sub-divide the set of qubits into two subsets, the even and the odd
qubits. For even (odd) qubits the sum of the coordinates of their
respective lattice site is even (odd).

Note that instead of with a bcc-symmetric lattice we
could have equivalently started with a cluster state on an sc-symmetric
lattice, because a cluster state on the latter is mapped to a 
cluster state on the former by $Z$-measurements on the qubits 
$(e,e,e)$ and $(o,o,o)$; see \cite{QCc}.

\begin{figure}
        \begin{center}
        \epsfig{file=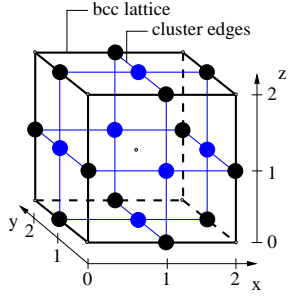, width=3.7cm}
        \caption{\label{cube}Elementary cell of the bcc
          lattice. Qubits live on the faces and edges of the
          elementary cell. Syndrome bits are located in the cube and
          on the sites. Each elementary cell has a volume of $2\times
          2 \times 2$ in cluster units. Black: elementary cell of the
          bcc lattice, blue: edges of the cluster graph.}        
        \end{center}
\end{figure}

\paragraph{The $[15,3,1]$ quantum Reed-Muller code.}By this we denote
a 15 qubit CSS-code based on the (classical) punctured Reed-Muller
code ${\cal{R}}(1,4)^*$ \cite{MacWilliams}. Its stabilizer generator
matrix has the form
\begin{equation}
  G_{RM} = \left(\begin{array}{c|c} G_X & 0 \\ 0 & G_Z \end{array}\right), 
\end{equation}
where 
\begin{equation}
  G_X = \left( 
  \begin{array}{lllllllllllllll} 
    1 & 0 & 1 & 0 & 1 & 0 & 1 & 0 & 1 & 0 & 1 & 0 & 1 & 0 & 1\\
    0 & 1 & 1 & 0 & 0 & 1 & 1 & 0 & 0 & 1 & 1 & 0 & 0 & 1 & 1\\
    0 & 0 & 0 & 1 & 1 & 1 & 1 & 0 & 0 & 0 & 0 & 1 & 1 & 1 & 1\\
    0 & 0 & 0 & 0 & 0 & 0 & 0 & 1 & 1 & 1 & 1 & 1 & 1 & 1 & 1
  \end{array}\right),
\end{equation}
and $G_Z$ is given by ${G_X}^{\perp}=G_Z \oplus (1,1,...,1,1)$.

This code has the fairly rare property that the encoded non-Clifford gate
$\overline{U}_z(\pi/4)= \exp(-i \pi/8 \overline{Z})$ is local
\cite{Kn,magic}, i.e.,
\begin{equation}
  \label{trans}
   \exp\left(-i \frac{\pi}{8}\, \overline{Z} \right) \cong
   \bigotimes_{i=1}^{15}  \exp\left(i \frac{\pi}{8}\, Z_i\right). 
\end{equation} 
This property has been used in magic state distillation \cite{magic}.
In the computational scheme described here we use it to fault-tolerantly
measure the encoded observables $\frac{\overline{X} \pm
  \overline{Y}}{\sqrt{2}}$ via {\em{local}} measurements of observables   
$\frac{X_i \pm Y_i}{\sqrt{2}}$.

\paragraph{Surface codes.}For the surface codes \cite{Kit1,Kit2} physical
qubits live on the edges of a two-dimensional lattice. The support of
a physical error must stretch across a constant fraction (typically 1/2)
of the lattice to cause a logical error. The protection against errors is
topological.  

The stabilizer generators of the code are associated
with the faces $f$ and the vertices $v$ of the lattice,
\begin{equation}
  S_X(v) = \bigotimes_{e|\, v \in \{\partial e\}}  X_e,\;\;
  S_Z(f)=\bigotimes_{e \in \{\partial f\}} Z_e.
\end{equation}  
Therein, $\partial$ is the boundary operator. The number of
qubits that can be stored depends on the boundary conditions of the
code lattice. The code
resulting from periodic boundary conditions, the `toric code' \cite{Kit1}, can
store two qubits. 

\begin{figure}
  \begin{center}
    \begin{tabular}{ll}
      a) & b)\\
      \parbox{7cm}{\epsfig{width=5.3cm, file=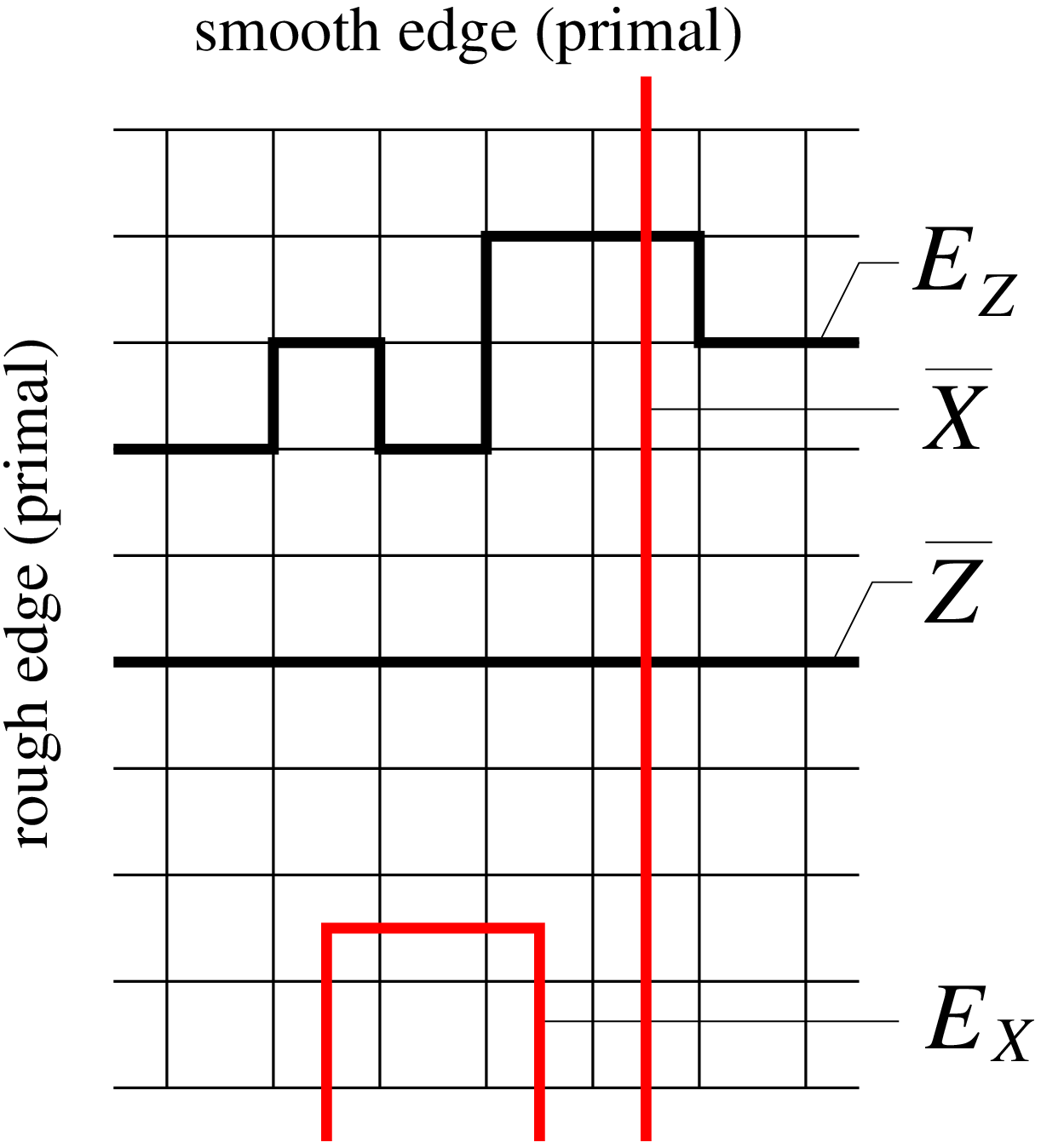}} &
      \hspace*{3mm}
      \parbox{5.3cm}{\epsfig{width=5.3cm, file=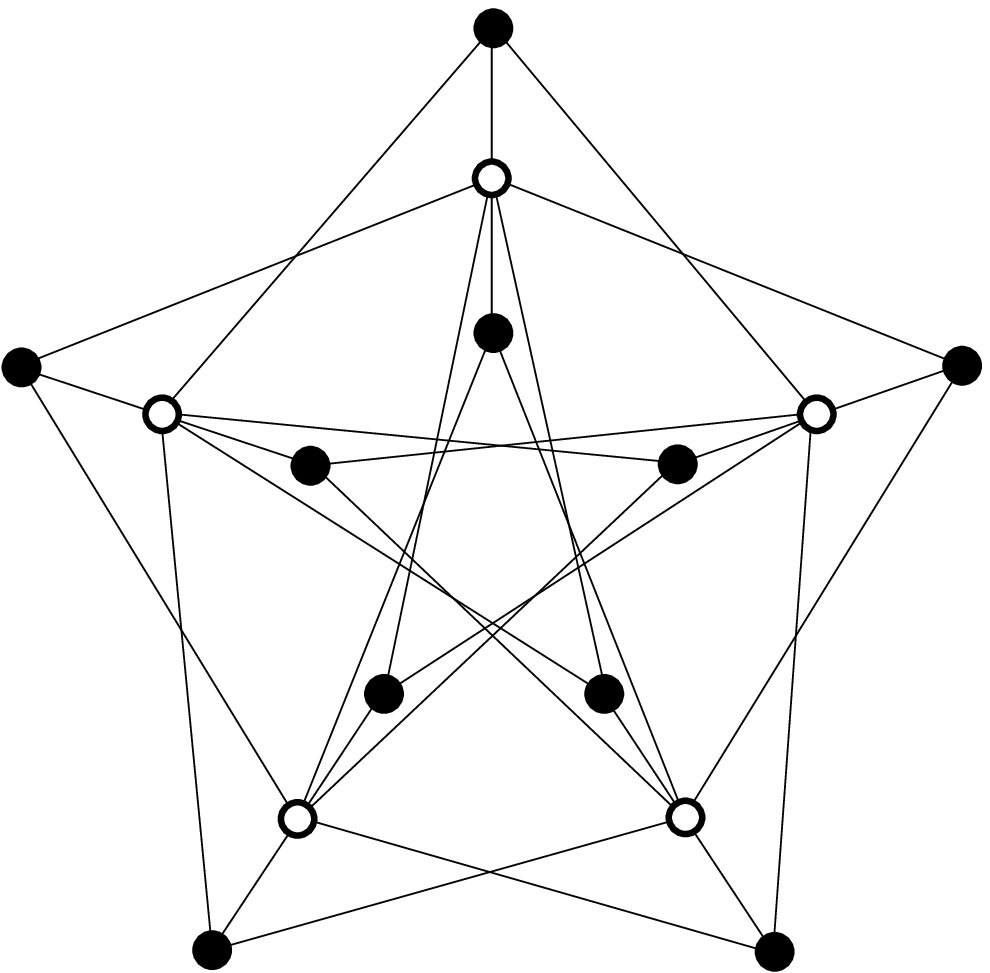}} 
    \end{tabular}
    \caption{\label{codes}Codes. a) Planar code. The encoded Pauli
      observables $\overline{X}$ and $\overline{Z}$, two errors and
      the different boundary types `smooth' and `rough' are shown. The
      errors $E_X$, $E_Z$ end in their respective rough
      boundary. $E_X$ corresponds to a trivial 1-cycle on  
      the dual lattice and has no effect on the encoded state. $E_Z$
      corresponds to an non-trivial 1-cycle on 
      the primal lattice. It leaves no syndrome but causes a logical
      $Z$-error. b) Bi-colorable graph state locally unitary
      equivalent to the $\overline{|+\rangle}$-state of the 15-qubit quantum
      Reed-Muller code.}
  \end{center}
\end{figure}

As an example we would briefly like to discuss the planar code
\cite{Kit2} which encodes one qubit; see Fig.~\ref{codes}a. 
This example exhibits many features
of our subsequent constructions one dimension higher up: Errors are
identified with 1-chains and show a syndrome only at their end
points. Homologically equivalent chains correspond to physically
equivalent errors. Error chains can end in the system boundary without
leaving a syndrome. 

Specifically, Pauli operators
$Z_i$ live on the edges of the primal (=shown) lattice, and
Pauli operators $X_j$ live on edges of the dual lattice. The encoded
$Z$-operator is a tensor product of individual $Z_i$ operators
corresponding to a primal 1-chain stretching from left to right
across the code lattice. The encoded $X$-operator corresponds to a
1-chain of the dual lattice that stretches from top to bottom. 

The code stabilizer is modified at the system boundary. For
example, a face to the left or right of the lattice has only three
elementary 1-chains in its boundary, instead of four. Such boundary is
called a `rough edge'. Where no modification of the faces occurs the system
boundary is a `smooth edge'. `Smooth' on the primal lattice is
`rough' on the dual, and vice versa. Error chains can end in a rough
edge of their respective lattice without leaving a syndrome, but not in
a smooth edge.   

The surface codes will occur rather implicitly in
our constructions. The reason is that here we do not use such codes to
encode logical  
qubits. Instead, we use them to appropriately ``wire'' a subset of the
cluster qubits, the $S$-qubits.
The link between surface codes and cluster states has been established
in \cite{LRE}, for the purpose of creating long range entanglement in
noisy 3D cluster states via local measurements. It has been found that the
error correction implemented by the local measurements is
described---like fault-tolerant data storage with the toric code---by
the so-called {\em{Random plaquette $\mathbb{Z}_2$-gauge model in
    three dimensions}} (RPGM) \cite{RPGM}. The
three-dimensional cluster state is like a surface code, one dimension
higher up. The third dimension, which is temporal in data storage
with the toric code, is spatial for the cluster state. The extra
spatial dimension can be used to fault-tolerantly mediate interaction
among qubits. The creation of an encoded Bell state over large
distances \cite{LRE} is the simplest example. The long-range quantum
correlations are engineered by the suitable choice of boundary
conditions.\medskip

Why are the  above two codes chosen?
For the fault-tolerant scheme of quantum computation described in this
paper we require a quantum code with the following three properties: 1) The
code is of CSS-type, 2) The code satisfies Eq.~(\ref{trans}), and
3) The code fits with cluster states. For some arrangement of qubits
on a translation-invariant two-dimensional lattice,
the code has a translation-invariant set of stabilizer generators and
these generators each have a small support on the lattice. 

The Reed-Muller code has properties 1 and 2 but not 3. The surface
codes have properties 1 and 3 but not 2. Thus, neither of the codes
alone suffices. But their combination does, as is described in the
subsequent sections.

\section{The measurement pattern}
\label{MeaP}

As pointed out in the introduction, we subdivide the cluster
$C$ into the three disjoint subsets $V$, $D$ and $S$. $S$ is
the the set of qubits where the non-Clifford part of the quantum
computation is performed, and $V$ and $D$ are to connect the qubits of
$S$ in the proper way. We have not yet explained what the defects
are, and we will do so only in the next section. For the moment it
suffices to note that the defects are located on the subset $D$ of the
cluster, and that $D$ is the union of the two disjoint
subsets $D_1$ and $D_2$. 
The measurement pattern on $D$, $S$ and $V$ is given by
\begin{equation}
  \label{MP}
  \begin{array}{lll}
    \mbox{Defect qubits}\,\, a \in D:& \mbox{Measurement of }\left\{
    \begin{array}{r} X_a \\ Z_a \end{array} \right.,& \mbox{if} \,\,
    \left\{
    \begin{array}{r} a \in D_2 \\ a \in D_1 \end{array}
    \right.,\vspace{1mm}\\  
    \mbox{Singular qubits}\,\, a \in S: & \mbox{Measurement of }
    \displaystyle{\frac{X_a\pm Y_a}{\sqrt{2}}},\vspace{1mm}\\
    \mbox{Vacuum qubits}\,\, a \in V: & \mbox{Measurement of}\,\, X_a.
  \end{array}
\end{equation}
Now we have to explain why we choose this measurement pattern, which
is best done using the language of homology.

\section{Involving the Reed-Muller quantum code}
\label{RM}

In this section we explain
the role of the Reed-Muller code for the described computational
scheme. Consider a cluster state $|\phi\rangle_{{\cal{C}}_2}$ on a
two-dimensional cluster ${\cal{C}}_2$. It is a resource for universal
quantum computation by measurements of the local observables $X$, $Z$
and $\frac{X\pm Y}{\sqrt{2}}$, \cite{QCc}. Denote by $Q \subset
{\cal{C}}_2$ the set of qubits which are measured in the eigenbasis of
$\frac{X\pm Y}{\sqrt{2}}$. These measurements implement the
non-trivial part of a quantum circuit. The measurements of
$X$ and $Z$ on the qubits  ${\cal{C}}_2\backslash Q$ implement the
Clifford part. They are performed simultaneously in the
first round of measurements. $|\Psi_\text{algo}\rangle_Q$ is the state
of the unmeasured qubits after the first measurement round. It is an
algorithm-specific stabilizer state, hence the 
subscript ``algo''. Since it is a stabilizer state, it is easy to
create and one may start with this state as an algorithm-specific
resource instead of the universal cluster state. Quantum computation
with this state proceeds by measuring local observables $\frac{X\pm
  Y}{\sqrt{2}}$. 

Now suppose an encoded version of this state,
$|\overline{\Psi}_\text{algo}\rangle_{S}$, was given. The state
were not perfect but only affected by local noise on the bare
level. Of course, a
question that arises immediately is how such a state is
obtained. The main part of work in this paper goes into answering this
question, see subsequent sections. Now, with
$|\overline{\Psi}_\text{algo}\rangle_{S}$ given, one could perform
fault-tolerant quantum computation by fault-tolerant measurement of
the encoded observables $\frac{\overline{X}\pm
  \overline{Y}}{\sqrt{2}}$. This is not what we have in mind, because
we are seeking a scheme of fault-tolerant quantum computation by
{\em{local}} measurements. Here lies the reason for involving the
(concatenated) Reed-Muller code: For this code, the fault-tolerant
measurement of the observables $\frac{\overline{X}\pm
  \overline{Y}}{\sqrt{2}}$ proceeds by local measurements of
observables $\frac{X\pm Y}{\sqrt{2}}$. The reason for this is property
(\ref{trans}). If $J$ is a set such that $g_X(J):=\bigotimes_{j \in J}X_j$ is
in the RM code stabilizer, then also
\begin{equation}
  g_\pm(J):=\bigotimes_{j \in J}\frac{X_j\pm
  Y_j}{\sqrt{2}} \in \mbox{RM code stabilizer}.
\end{equation}
The relevant encoded observables are given by
\begin{equation}
  \frac{\overline{X}\pm
  \overline{Y}}{\sqrt{2}} = \bigotimes_j \frac{X_j \mp
  Y_j}{\sqrt{2}}.
\end{equation}
The ``$\mp$'' is for the total number of concatenation levels being odd. If
the number is even, replace ``$\mp$'' by ``$\pm$''. Therefore, if all the
bare qubits belonging to an encoded qubit are individually measured in
the eigenbasis of $\frac{X_j-Y_j}{\sqrt{2}}$, then the eigenvalue found
in a measurement of the encoded observable $\frac{\overline{X}+
  \overline{Y}}{\sqrt{2}}$ and the eigenvalues of 
the stabilizer generators $g_+(J)$ can be deduced from the individual
measurement outcomes. This is all what is needed for fault-tolerant
measurement of the encoded observable $\frac{\overline{X}+
  \overline{Y}}{\sqrt{2}}$.

The $Z$-part of the code stabilizer is lost in the local measurement,
but it is not needed for the fault-tolerant measurement of
$\frac{\overline{X}\pm \overline{Y}}{\sqrt{2}}$. This can be seen as
follows. For simplicity assume a local depolarizing error $p/3
\left([X_j]+[Y_j]+[Z_j]\right) = \frac{p}{3 \sqrt{2}}
\left( \left[X_j+Y_j \right]
+ \left[X_j-Y_j\right] + \sqrt{2}[Z_j] \right)$ for all
qubits $j$. The brackets ``$[\cdot]$'' indicate a
super-operator. W.l.o.g. assume that the local measurements are in the
eigenbasis of $\frac{X_j-Y_j}{\sqrt{2}}$. Then, the error
$\frac{X_j-Y_j}{\sqrt{2}}$ is absorbed in the measurement and has no
effect. The second error $\frac{X_j+Y_j}{\sqrt{2}} =
i  \frac{X_j-Y_j}{\sqrt{2}} Z_j \cong Z_j$, so all remaining errors
are equivalent to $Z$-errors. Such errors are identified by the stabilizer
elements $\{g_+(J)\}$.   

To summarize, the Reed-Muller code is involved to perform the
fault-tolerant measurement of the encoded observables
locally on the bare level. The eigenvalues corresponding to the
encoded observable and to the relevant 
stabilizer generators are simultaneously inferred from the measurement
outcomes. Error correction proceeds by classical post-processing of
these quantities.

\section{Involving a topological quantum code - Homology}
\label{homology}

The remaining question is how we actually create the state
$|\overline{\Psi}_\text{algo}\rangle_{S}$ with only local
error from a three-dimensional cluster state. To accomplish this task
we involve topological error correction.  

\subsection{Errors and correlations as chains}
\label{ecc}

The physical objects of our discussion---cluster state
correlations and error operators---may be identified with faces and
edges of an underlying lattice. Compositions of such
faces or edges are called 2-chains and 1-chains, respectively. For the
chains homology provides an equivalence relation; namely, two chains
are homologically equivalent if they differ by the boundary of a third
chain one dimension higher up \cite{Vick, Hatcher, Nakahara}. Homology
plays a role in our constructions  
because homological equivalence of the underlying chains implies
physical equivalence of
the associated physical objects. 

First we introduce the two simple cubic
sub-lattices ${\cal{L}}$ and
$\overline{\cal{L}}$ whose vertices are at locations 
\begin{equation}
  \begin{array}{rcl}
    {\cal{L}}&:&\{(e,e,e)\},\\
    \overline{\cal{L}}&:&\{(o,o,o)\}.
  \end{array}
\end{equation} 
One lattice can be obtained from the other via translation by a vector
$(\pm 1,\pm 1,\pm 1)$. ${\cal{L}}$ and $\overline{\cal{L}}$ are dual
to another in the sense that the faces of ${\cal{L}}$ are the edges of 
$\overline{\cal{L}}$, the cubes of ${\cal{L}}$ are the 
vertices of $\overline{\cal{L}}$, and vice versa. We denote by $*$ the
duality transformation that maps primal edges into the corresponding
dual faces ($*e=\overline{f}$), and so forth.

We denote by ${\cal{B}}(C_0):=\{v_k\}$ the set of vertices in
  ${\cal{L}}$, by ${\cal{B}}(C_1) =\{e_l\}$ the set of edges in 
${\cal{L}}$, by ${\cal{B}}(C_2)=\{f_m\}$ 
the set of faces in ${\cal{L}}$, and by ${\cal{B}}(C_3) =\{q_n\}$ the
set of elementary cells [or cubes] of ${\cal{L}}$. We may now define
  chains in ${\cal{L}}$ \cite{Vick}. ${\cal{B}}(C_0)$
 forms a basis for the set $C_0$ of so called 0-chains $c_0$, ${\cal{B}}(C_1)$
 forms a basis for the set $C_1$ of 1-chains $c_1$, and so
  forth. Specifically, the chains are given by 
\begin{equation}
  \begin{array}{rclcrclcrclcrcl}
    c_0&=&\sum_k z_k v_k,\;\; &  c_1&=&\sum_l z_l e_l,\;\;&  c_2&=&\sum_m z_m
    f_m, \;\;&  c_3&=&\sum_n z_n q_n.
  \end{array}
\end{equation}
where $z_k,z_l,z_m,z_n \in \mathbb{Z}_2$. The sets $C_0$, $C_1$, $C_2$
and $C_2$ are, in fact, abelian groups under component-wise addition,
e.g. $c_1+c_1^\prime=\sum_lz_le_l+\sum_lz_l^\prime e_l=
\sum_l(z_l+z_l^\prime) e_l$. For each $i=1..3$, there exists a
homomorphism $\partial_i$ mapping $C_i$ to $C_{i-1}$, with the
composition $\partial_{i-1} \circ \partial_i=0$. Then,
\begin{equation}
  {\cal{L}}=\{C_3, C_2, C_1, C_0\}
\end{equation} 
is called a chain complex, and $\partial$ is called boundary
operator. It maps an $i$-chain $c_i$ to its boundary, which is an
$i-1$-chain. In the same way, $\overline{\cal{L}}$ can be defined as a
dual chain complex,
$\overline{\cal{L}}=\{\overline{C}_3,\overline{C}_2, \overline{C}_1,
  \overline{C}_0\}$, with chains $\overline{c}_3$, $\overline{c}_2$,
  $\overline{c}_1$ and $\overline{c}_0$.    

Now, considering a space ${\cal{C}}$, two chains $c_n,\, c_n^\prime \in
C_n({\cal{C}})$ are homologically equivalent if $c_n^\prime=c_n +
\partial c_{n+1}$ for some $c_{n+1} \in C_{n+1}({\cal{C}})$
\cite{Vick}. Of interest for topological error-correction is the
notion of {\em{relative homology}}. Consider a pair of spaces
$({\cal{C}}, D)$ with $D\subset {\cal{C}}$. Then, two chains $c_n$,
$c_n^\prime$ are called equivalent w.r.t relative homology, $c_n^\prime
\cong_\text{r} c_n$, if
$c_n^\prime = c_n+ \partial c_{n+1}+ \gamma_n$ for some $c_{n+1} \in
C_{n+1}({\cal{C}})$, $\gamma_n \in C_n(D)$; see \cite{Hatcher}. Relative
  cycles may end in $D$.

Below we describe how the cluster state quantum correlations may be
identified with the 2-chains, the errors with the 
1-chains and the syndrome with the 0-chains of ${\cal{L}}$ and
$\overline{\cal{L}}$. All these objects appear in two kinds, `primal' and
`dual', depending on whether they are defined with respect to ${\cal{L}}$ or
$\overline{\cal{L}}$.

\paragraph{Cluster state correlations.}We define primal such correlations,
$K(c_2)$, and dual ones, $K(\overline{c}_2)$, which can be
identified with 2-chains in ${\cal{L}}$ and $\overline{\cal{L}}$,
respectively, by 
\begin{equation}
  \label{Hcorr}
  K(c_2):=\prod_{f \in \{c_2\}} K_f,\;\;  
   K({\overline{c}_2}):=\prod_{\overline{f} \in \{\overline{c}_2\}} 
   K_{\overline{f}}. 
\end{equation}
Therein, e.g. the set $\{c_2\}$ is defined via a mapping
$c_2=\sum_mz_mf_m \longrightarrow \{c_2\}=\{f_m|z_m=1\}$. Further, we
introduce the notion
$O(c):=\bigotimes_{a \in\{c\}} O_a$, for all $c \in C$ and $O \in \{X,Z\}$.
It is now easily verified that
\begin{equation}
  \label{Hcorr2}
  K(c_2)=X(c_2)Z(\partial c_2),\; K(\overline{c}_2)=X(\overline{c}_2)Z(\partial
  \overline{c}_2).   
\end{equation}

\paragraph{Errors.}We will mainly discuss (correlated) probabilistic
noise. Then, it is sufficient to restrict the attention to
Pauli phase flips $Z$, because $X_a \cong \bigotimes_{b\in N(a)} Z_b$
etc.  We combine $Z$-errors on
odd (even) qubits to primal (dual) error chains $E(c_1)$
($E(\overline{c}_1)$),
\begin{equation}
  E(c_1):=Z(c_1),\;\;  E(\overline{c}_1):= Z(\overline{c}_1).  
\end{equation}

\paragraph{Syndrome.}The first type of correlations we discuss are
those for error 
correction in $V$. They are characterized by the property that the
corresponding 2-chains have no boundary. I.e., we consider $K(c_2)$,
$K(\overline{c}_2)$ with $\{c_2\}, \{\overline{c}_2\} \in V$ and
$\partial c_2 = \partial \overline{c}_2 = 0$. With (\ref{Hcorr}), these
correlations take the form $K(c_2)=X(c_2)$, 
$K(\overline{c}_2)=X(\overline{c}_2)$. They are 
measured by the $X$-measurements in $V$, see 
(\ref{MP}), and are used to identify errors occurring on
the qubits $\{c_2\}$, $\{\overline{c}_2\}$.

The group of 2-chains in the kernel of $\partial$ we denote by
$Z_2({\cal{L}})$. Since $\partial\partial =0$, a subgroup of
those, $B_2({\cal{L}})$, is 
formed by the 2-chains which are themselves a boundary of a 3-chain.
Denote by $q$ ($\overline{q}$) a 3-chain from the basis
${\cal{B}}(C_3)$ (${\cal{B}}(\overline{C}_3)$). It represents 
an individual cell [or cube] of the primal lattice ${\cal{L}}$ (dual
lattice $\overline{\cal{L}}$). The associated quantum correlations are
\begin{equation}
  \label{TopECC}
  K_q:=K(\partial q)= X(\partial q),\;\; K_{\overline{q}}:=K(\partial
 \overline{q})= X(\partial \overline{q}). 
\end{equation}
When being measured, each of these correlations yields a syndrome bit $Sy(q)$,
$Sy(\overline{q})$ which, we say, is located at $q$ or $\overline{q}$,
respectively. Because the lattices ${\cal{L}}$ and
$\overline{\cal{L}}$ are dual to 
another, we may identify the cell $q$ in ${\cal{L}}$ with a vertex
$\overline{v}$ in $\overline{\cal{L}}$, and vice versa. In this way,
the syndrome 
bits become located at vertices of the lattices ${\cal{L}}$ and
$\overline{\cal{L}}$. The 
syndrome resulting from the quantum correlations (\ref{TopECC}) are those
which enable topological error correction \cite{RPGM}.

\paragraph{Syndrome and errors.} Let $E(c_1)$ denote a primal
error chain, $\overline{q} \in {\cal{B}}(\overline{C}_3)$ a cell in
the dual lattice $\overline{\cal{L}}$ and 
$v \in {\cal{B}}(C_0)=*\overline{q}$.  
$K(\overline{q})$ detects the error $E(c_1)$ if
$|\{c_1\} \cap \{\partial \overline{q}\}| = \mbox{odd}$. Equivalently,
$K(\overline{q})$ detects $E(c_1)$ if $v \in
\{\partial c_1\}$. Thus, error chains show a syndrome only at their
ends.

\paragraph{Correlations and errors.}Primal cluster state correlations
are affected by dual error chains and dual correlations are affected
by primal error chains. Primal correlations are not affected by primal
error chains, and dual correlations are not affected by dual error
chains. 

To see this, note that a primal correlation $K(c_2)$ consists of Pauli
operators $X$ on even qubits and Pauli operators $Z$ on odd qubits. A
dual error chain $E(\overline{c}_1)$ consists of operators $Z$ on
even qubits.  Then, $E(\overline{c}_1)
K(c_2) = (-1)^{|\{\overline{c}_1\} \cup \{c_2\}|}\, K(c_2)
E(\overline{c}_1)$.  If $|\{\overline{c}_1\} \cup \{c_2\}|$ is odd,
the correlation 
$K(c_2)$ is conjugated to $-K(c_2)$ by the error. If it is even, then
the correlation remains unchanged.  This situation has a geometric
interpretation. $|\{\overline{c}_1\} \cup \{c_2\}|$ is the number
of intersection points between the primal 2-chain $c_2$ and the dual
1-chain $\overline{c}_1$. If the number of intersections is odd (even)
then the correlation is (is not) affected by the error. For dual
correlations and primal errors the situation is the same.
 Further, a primal error chain consists of Pauli operators $Z$ on odd
qubits. Thus, $[K(c_2),E(c_1)]=0$ always. Similarly, 
$[K(\overline{c}_2),E(\overline{c}_1)]=0$, $\forall\, \overline{c}_1,
  \overline{c}_2$. 

\paragraph{Defects.}The purpose of defects is to structure the space
underlying the pair of lattices ${\cal{L}}, \overline{\cal{L}}$.
Practically, a defect can be thought of as a set of qubits that are
removed from the initial cluster 
${\cal{C}}$ before the remaining qubits are entangled. For the chain
complexes $C$, $\overline{C}$, a defect is a set $d$ of missing
edges. What defines a defect as an entity is that the belonging edges
are connected. As all the other objects, defects are either primal or dual,
\begin{equation}
  \label{defdef}
  d \subset {\cal{B}}(C_1),\; \overline{d} \subset {\cal{B}}(\overline{C}_1).
\end{equation}
The sets $d$, $\overline{d}$ of defect qubits are not
arbitrary. Seen from afar they take the shape of doughnuts. These
doughnut-shaped defects will be topologically entangled with another, and
the way they are entangled encodes the quantum algorithm to be
performed. From the viewpoint of quantum logic, what matters about the
doughnuts is that they are loops. Their `thickness' is required for
fault-tolerance. 

We now briefly explain how the above definition of a defect as a set of
missing cluster qubits fits with the measurement pattern
(\ref{MP}). Formally, each defect $d$ will be assigned a set
$D(d)$ of locations on the cluster. This set is subdivided into a set
of edge- and a set of face qubits, $D_1(d)$ and $D_2(d)$. 
Here, the notions of `edge' and `face' are in reference to the
lattice the defect belongs to. If the defect is primal (dual) then the
edges and faces are taken with respect to the primal (dual) lattice. 
For primal defects, the sets $D_1(d)$ and $D_2(d)$ are defined as
\begin{equation}
  D_1(d):=d,\; D_2(d)=\left\{f \in {\cal{B}}(C_2)|\, \{\partial f\} \cap
  d= \{\partial f\} \right\}.
\end{equation} 
For dual defects, replace $f$ by $\overline{f}$ and $C_2$ by
$\overline{C}_2$ in the above definition. The whole defect region $D$
splits into an edge part $D_1$ and a face part $D_2$, $D=D_1\cup D_2$
where
\begin{equation}
  D_1 = \bigcup_d D_1(d),\; D_2=\bigcup_d D_2(d).
\end{equation}
Now the measurement pattern (\ref{MP}) becomes understandable: the
edge qubits in the defects are measured in the $Z$-basis which
effectively removes them from the cluster \cite{BR}. In this way, the
quantum state on the exterior of the defect becomes disentangled from
the state with support on interior of the defect. Thereby, a defect is
created in the cluster lattice. Note that the qubits on faces whose
entire boundary is in $D_1(d)$ become disentangled individually. If no
errors were present we could leave these qubits alone. However, their
measurement in the $X$-basis provides additional syndrome and so it
is advantageous to measure them.

\paragraph{Correlations and defects.}

In the proximity of a primal defect, edges in the boundary of a primal 2-chain
are removed, see (\ref{defdef}). Therefore, primal correlations can
end in primal 
defects. Dual defects do not remove primal edges, and thus primal
correlations cannot end in dual defects. Analogously, dual
correlations can end in dual defects, but not in primal defects.

\paragraph{Syndrome and defects.}In the presence of a  primal
defect $d$ the correlations
$\{K_{\overline{q}}=K(\partial \overline{q})|\, \overline{q} \in
{\cal{B}}(\overline{c}_3) \, \wedge\, \partial \overline{q} \cap d \neq
  \emptyset \}$ do not commute with the measurements (\ref{MP}), such that the
  syndrome at the locations
\begin{equation}
  D_0=\{v \in \partial e|\, e \in d \}
\end{equation}
is lost. Note, however, that for each defect $d$ there will be one
syndrome bit associated with the defect as a whole. There exists a
2-cycle $\overline{c}_2(d)$, $\{\overline{c}_2(d)\} \subset V$, that
wraps around $d$, and
$K(\overline{c}_2(d))=X(\overline{c}_2(d))$. When the qubits in $V$
are measured in the $X$-basis, this correlation yields an additional
syndrome bit. Dual defects act analogously on the dual lattice.

\paragraph{Errors and defects.}

Because the local syndrome is
lost at the surface of a defect, primal error chains can potentially
end in primal defects. However, there is a dual correlation
$K(\overline{c}_2(d))=X(\overline{c}_2(d))$ wrapping
around a primal defect, and this correlation detects a primal error
chain $E(c_1)$ if the number of intersection points between
$\overline{c}_2(d)$ and $c_1$ is odd. Thus, primal error chains can
{\em{pairwise}} end in primal defects. 

Primal error chains cannot end in dual defects, because dual defects do not
remove primal syndrome. Similarly, dual error chains can pairwise end in
dual defects, and they cannot end in primal defects. 
\medskip

\noindent
The relations among cluster state correlations, errors and defects are
summarized in Tab.~\ref{relations}. 

\begin{table}
  \begin{center}
    \begin{tabular}{|l|l|l|l|l|l|} \hline
       & \multicolumn{5}{|c|}{\textbf{This object does ...}}\\
      \textbf{to this one}$\downarrow$ & dual corr. &
      primal defect & dual defect & primal err. 
      cy. & dual err. cy.\\ \hline
      correlation & nothing 
      & bound & repel & nothing & affect\\ \hline
      dual correlation & & repel & bound & affect & nothing\\ \hline
      primal defect & & & \parbox{2.5cm}{encircle} &
      \parbox{2.3cm}{pairwise end} & \parbox{2cm}{encircle}\\ \hline
      dual defect & & & & \parbox{2cm}{encircle} &
      \parbox{2.3cm}{pairwise end} \\ \hline
      primal err. cy. & & & & & nothing\\ \hline
    \end{tabular}
  \end{center}
  \caption{\label{relations}This table shows who does what to
      whom. `$A$ bounds $B$' is synonymous with `$B$ ends in $A$'. The
      displayed objects do not interact with themselves.} 
\end{table}

\subsection{Homological and physical equivalence}

We have so far identified physical objects---correlations and
errors---with chains of a chain complex. In this section we point out
that it
is the homology class of the chain rather than the chain itself which
characterizes the respective physical object. The equivalence of two
chains under relative homology implies the physical equivalence of the
corresponding physical operators. 

1. Cluster state correlations. We regard two cluster state
correlations $K(c_2)$, $K(c_2^\prime)$ as 
physically equivalent if they yield the same stabilizer element for the state
$|\overline{\Psi}_\text{algo}\rangle_S$ after the measurement of the
qubits in $V$ and $D$. This requires two things. First, $K(c_2)$ and 
$K(c_2^\prime)$ need to be simultaneously measurable. With $O_a$
the locally measured observables (\ref{MP}) we require $[K(c_2), O_a] =0\;
\forall a \in V\cup D \Longleftrightarrow [K(c_2^\prime ), O_a] =0\;
\forall a \in V\cup D$. Second, the two operators
must agree on $S$, $K(c_2)|_S = K(c_2^\prime)|_S$. Then, the following
statement holds: If $c_2^\prime \cong_\text{r} c_2$ w.r.t. $(V\cup
D, D)$ then 
$K(c_2^\prime) \cong K(c_2)$. \\ 
{\em{Proof}}: There exists $c_3 \in C_3 |\, \{\partial c_3\} \subset V
\cup D_2$ and $\gamma_2 \in C_2|\, \{\gamma_2\} \subset D_2$ such that
$c_2^\prime = c_2 + \partial c_3 + \gamma_2$. 1. Simultaneous
measurability: $K(\gamma_2) = \left( \bigotimes_{a \in  
  D_2(\gamma_2)}X_a \right) \left( \bigotimes_{b \in
  D_1(\gamma_2)}Z_b \right)$, where $D_1(\gamma_2) \subset D_1$ and
$D_2(\gamma_2) \subset D_2$. Therefore, with (\ref{MP}),
$[K(\gamma_2), O_a] =0\; 
\forall a \in V\cup D$ (*). Similarly, $K(\partial c_3)=X(\partial c_3)$ 
such that, with (\ref{MP}),  $[K(\partial c_3), O_a] =0\;
\forall a \in V\cup D$ (**). Since $K(c_2^\prime) = K(c_2) K(\partial
c_3) K(\gamma_2)$, (*) and (**) imply simultaneous
measurability of $K(c_2)$ and $K(c_2^\prime)$ on $V \cup D$. 2. Same
restriction to $S$: $K(\partial c_3)$ and $K(\gamma_2)$ don't act on
$S$, hence  $K(c_2)|_S = K(c_2^\prime)|_S$. $\Box$

2. Errors. Two errors $E(c_1)$ and $E(c_1^\prime)$ are 
physically equivalent if they cause the same
damage to the computation. That is, they have the same logical effect
and leave the same syndrome. Then, the following statement holds: If
$c_1^\prime \cong_r c_1$ w.r.t. $(V \cup D, D)$ then
$E(c_1^\prime) \cong E(c_1)$. \\
{\em{Proof}}:  There exist $c_2 \in C_2,\, 
\gamma_1 \in C_1,\,\, \mbox{with } \{c_2\} \subset V \cup D_2, \, \{\gamma_1\}
\subset D_1$, such that $c_1^\prime = c_1 + \partial
c_2 + \gamma_1$. Now, a Pauli spin flip error
$X$ is absorbed in a subsequent $X$-measurement and has no effect on
the computation, $\frac{I \pm X}{2} X = \pm \frac{I \pm X}{2}
I$. Thus, with (\ref{MP}), $X_a \cong I_a$ for all $a \in V \cup D_2$.
Similarly, $Z_b \cong I_b$ for all $b \in D_1$. Then, $E(c_1+\partial
c_2 + \gamma_1) = E(c_1) Z(\partial c_2) Z(\gamma_1) 
  = E(c_1) K(c_2) X(c_2) Z(\gamma_1) \cong E(c_1)$. $\Box$

\section{Constructive techniques}
\label{Elm}

The purpose of the measurements in $V$ and $D$ is to create on $S$ the
Reed-Muller-encoded algorithm-specific resource
$|\overline{\Psi}_\text{algo}\rangle_S$ described in
Section~\ref{RM}. In Section~\ref{Sloc}, we specify the location of
$S$-qubits with 
respect to the lattice defects and then, in
Section~\ref{BC},  we give a construction for a topologically protected
circuit providing $|\overline{\Psi}_\text{algo}\rangle_S$.  
 
\subsection{Location of the $S$-qubits}
\label{Sloc}

\begin{figure}
  \begin{center}
    \begin{tabular}{lclcl}
    a) && b) && c)\\ 
    \parbox[t]{3.5cm}{\epsfig{width=3.5cm, file=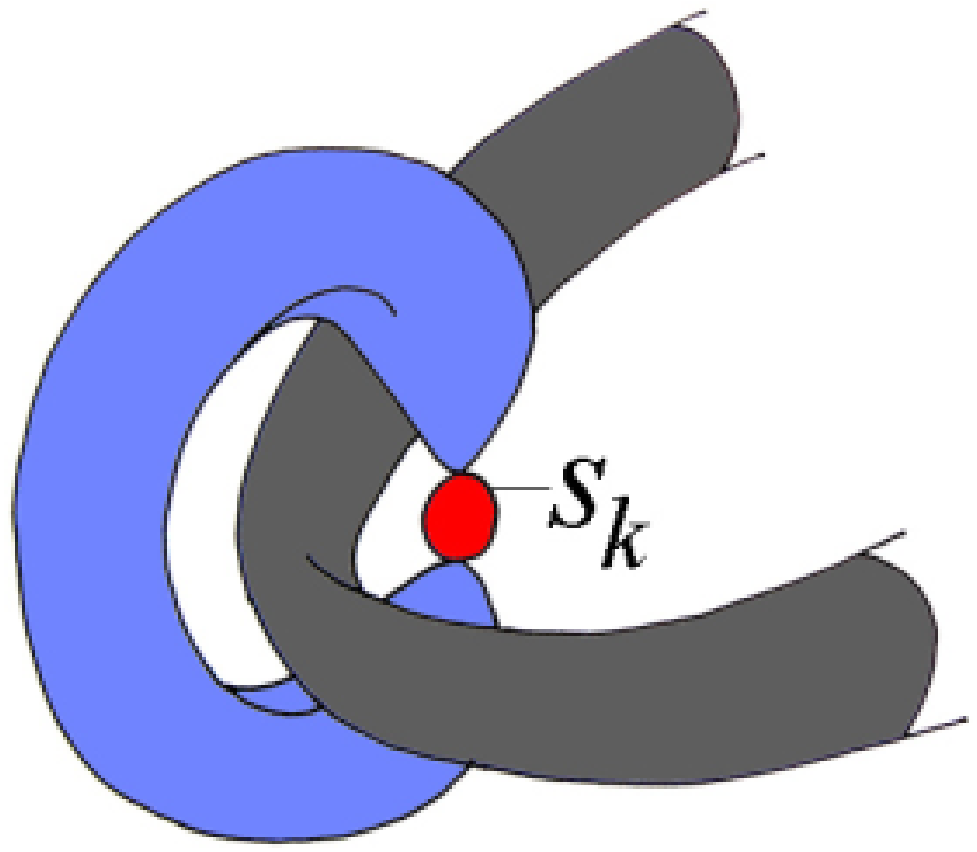}} &&
    \parbox[t]{3.5cm}{\epsfig{width=3.5cm, file=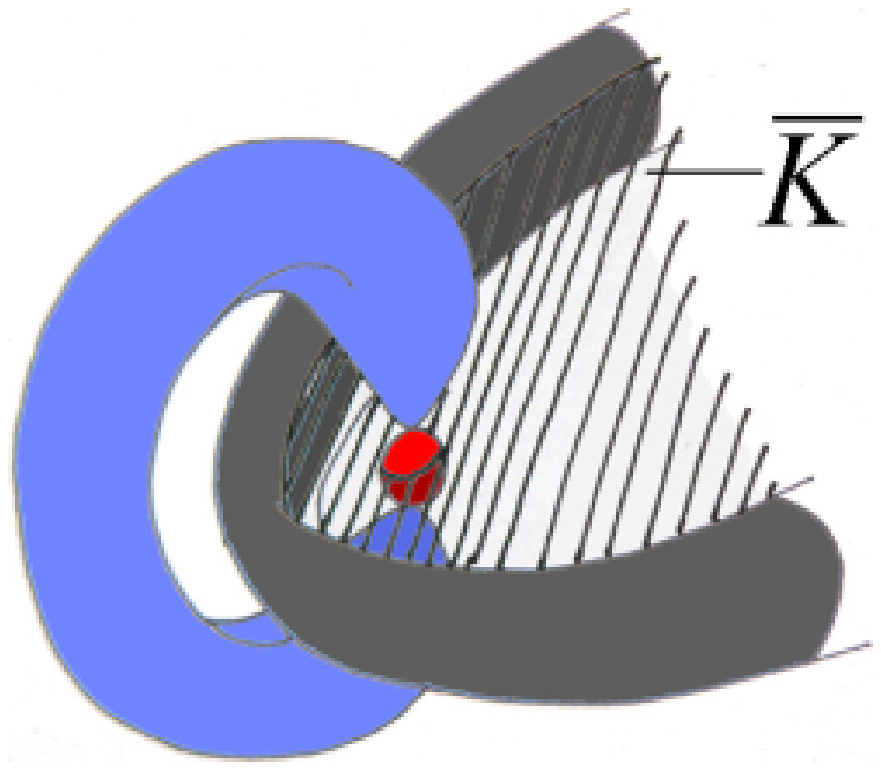}} &&
    \parbox[t]{3.5cm}{\epsfig{width=3.5cm, file=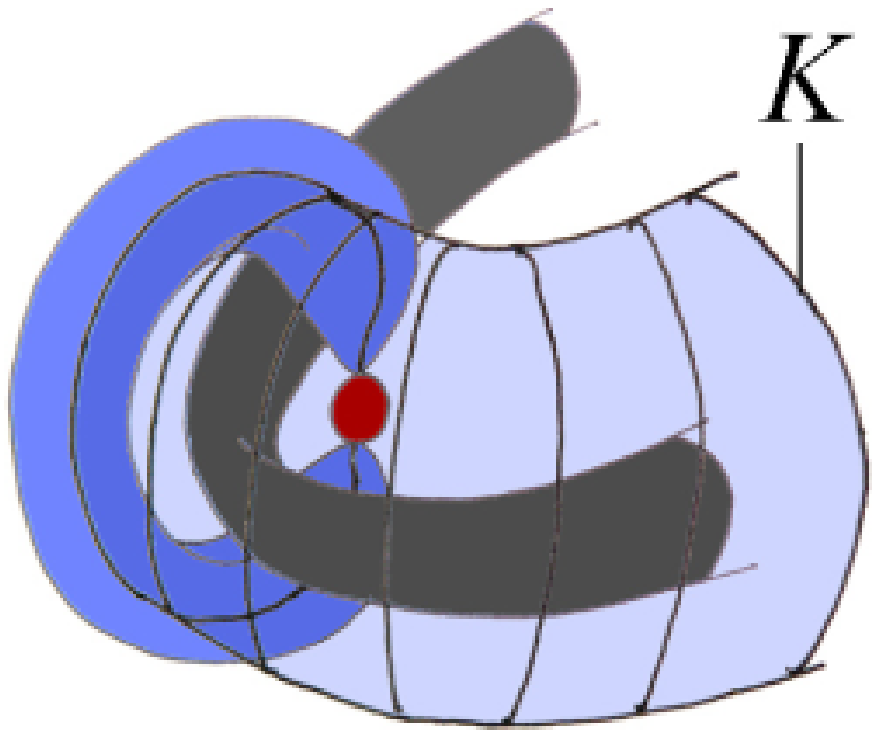}} 
    \end{tabular}
    \caption{\label{SLOC}a)  Location of a singular qubit (dual, held
    by a primal defect). b,c) The correlations $\overline{K}$,
    $K$  affect the $S$-qubit by a
    Pauli-operator $X$ in (b), and by $Z$ in (c).}
  \end{center}
\end{figure}

The  $S$-qubits have very particular locations within
the cluster. Besides the defects in the shape of doughnuts that we
have already introduced the cluster also supports defects shaped like
ear clips. The opening of these ear clip defects is only one cluster
qubit wide. If this one cluster qubit were a defect qubit ($q \in D$)
too, the ear clips 
would become doughnuts. But the particular cluster qubit is not in $D$, it
is an $S$-qubit. The situation is displayed graphically in
Fig.~\ref{SLOC}a.   

The appropriate stabilizer generators among the $S$-qubits are
induced  from the
cluster state correlations associated  with relative 2-cycles, by
measurement of the $V$- and $D$-qubits.
As everything else in this computational scheme, the $S$-qubits occur
in the two kinds `primal' and `dual'. We 
call an $S$-qubit $q$ primal, $q \in S_p$, if it lives on the a face
of the primal lattice, and we call it dual, $q \in S_d$,
if it lives on a face of the dual lattice.  

We now discuss how primal and dual correlations affect the
$S$-qubits. Consider, for example the correlation $K(c_2)$ corresponding to a
primal relative 2-chain $c_2$.  A primal $S$-qubit at
location $q$ may lie within  $c_2$, but never in
its boundary, $\{\partial c_2\} \cap q =
\emptyset$. A dual $S$-qubit $q^\prime$ may lie in the boundary of a
primal 2-chain $c_2$ but never in $c_2$ itself, $\{c_2\} \cap q^\prime
=\emptyset$. We therefore conclude that 
\begin{equation}
  \label{KaffeS}
  \mbox{\parbox{14cm}{\em{
	A primal correlation $K(c_2)$ acts on a
	primal $S$-qubit by one of the two Pauli-operators $X$, $I$ and on a
	dual $S$-qubit by one of the two operators $Z$, $I$. A 
	dual correlation $K(\overline{c}_2)$ acts on a
	primal $S$-qubit by one of the two Pauli-operators $Z$, $I$ and on a
	dual $S$-qubit by one of the two operators $X$, $I$.}}}
\end{equation}
For finding the extended relative 2-cycles on the primal lattice, we thus
regard $S_p$ as part of $V$ and $S_d$ as part of $D$. Analogously, for
finding the extended relative 2-cycles on the dual lattice, we 
regard $S_p$ as part of $D$ and $S_d$ as part of $V$.
In this way, the problem of finding the extended primal 2-cycles on a
cluster with $S$-qubits is reduced to the same problem
without $S$-qubits.

\subsection{Creating $|\overline{\Psi}_\text{algo}\rangle$ among the
  $S$-qubits} 
\label{BC}

The construction of a topologically protected circuit providing
$|\overline{\Psi}_\text{algo}\rangle_S$ proceeds in three steps. First
we show that $|\overline{\Psi}_\text{algo}\rangle_S$ is local
unitarily equivalent to a bi-colorable graph state, by local
Hadamard-transformations. Second, we show how to create a bi-colorable
graph state. Third, we take care of the Hadamard-transformations.

\paragraph{Equivalence of $|\overline{\Psi}_\text{algo}\rangle$ to a
  bi-colorable graph state.}   Denote by
$|\Psi\rangle_{{\cal{C}}_2}$ the state after the first measurement
round ($X$- and $Z$-measurements) in a
\QCcns-computation on a cluster state $|\phi\rangle_{{\cal{C}}_2}$,
cf. Section~\ref{RM}. Denote by ${\cal{C}}_X$, ${\cal{C}}_Z$ the
subsets of ${\cal{C}}_2\backslash Q$ whose qubits are measured in the
$X$- or $Z$-basis, respectively. Further, denote by
${\cal{C}}_\text{even}$, ${\cal{C}}_\text{odd}$ the sets of even and
odd qubits in ${\cal{C}}_2$ (checkerboard pattern).
The state $|\Psi\rangle_{{\cal{C}}_2}$ is given
by 
\begin{equation}
  \label{DePsi}
  |\Psi\rangle_{{\cal{C}}_2} \sim \left(\bigotimes_{a \in {\cal{C}}_X}
   \frac{I \pm X_a}{2} \right) \left(\bigotimes_{b \in {\cal{C}}_Z}
   \frac{I \pm Z_b}{2} \right) |\phi\rangle_{{\cal{C}}_2}
\end{equation}
and has the form
$|\Psi\rangle_{{\cal{C}}_2}=|\Psi_\text{algo}\rangle_Q \otimes
|\mbox{rest}\rangle_{{\cal{C}}_2\backslash Q}$. Further,
$|\phi\rangle_{{\cal{C}}_2}$ is local unitary equivalent to some
CSS-state, $|\phi\rangle_{{\cal{C}}_2}= \left(\bigotimes_{i \in
  {\cal{C}}_\text{odd}} H_i \right) |CSS\rangle_{{\cal{C}}_2}$. Then,
with (\ref{DePsi}) and ${\cal{C}}_X^\prime:=\left({\cal{C}}_X \cap
{\cal{C}}_\text{even}\right) \cup \left({\cal{C}}_Z \cap
{\cal{C}}_\text{odd}\right)$,  ${\cal{C}}_Z^\prime:=\left({\cal{C}}_Z \cap
{\cal{C}}_\text{even}\right) \cup \left({\cal{C}}_X \cap
{\cal{C}}_\text{odd}\right)$,
$
    |\Psi\rangle_{{\cal{C}}_2} \sim  \left(\bigotimes_{i \in
  {\cal{C}}_\text{odd}} H_i \right) \left(\bigotimes_{a \in
  {\cal{C}}_X^\prime}
   \frac{I \pm X_a}{2} \circ \bigotimes_{b \in {\cal{C}}_Z^\prime}
   \frac{I \pm Z_b}{2} |CSS\rangle_{{\cal{C}}_2} \right)
   = \left(\bigotimes_{i \in
  {\cal{C}}_\text{odd}} H_i \right) |CSS^\prime\rangle_{{\cal{C}}_2}.
$
Thus, also the state $|\Psi_\text{algo}\rangle$ is l.u.
equivalent to a CSS-state, 
\begin{equation}
|\Psi_\text{algo}\rangle_Q =
\left(\bigotimes_{q \in {\cal{C}}_\text{odd} \cap Q} H_q \right)
|CSS^{\prime\prime}\rangle_Q.
\end{equation}  
Now, we consider the concatenated-Reed-Muller-encoded
resource $|\overline{\Psi}_\text{algo}\rangle_S$,
which may obtained from the  bare state
$|\Psi_\text{algo}\rangle_Q$ via encoding, 
$|\overline{\Psi}_\text{algo}\rangle_S = \left(\bigotimes_{q \in
  Q}\mbox{Enc}_q \right) |\Psi_\text{algo}\rangle_Q$.
The encoding procedure $\mbox{Enc}$ takes every qubit $q \in Q$ to a
set $S(q)$ of 
qubits, $\bigcup_{q \in Q} S(q) = S$. It has the property that 
$\mbox{Enc}_q \circ H_q = \left(\bigotimes_{i \in S(q)} H_i \right)
  \circ \mbox{Enc}^\prime_q$.
Therein, $\mbox{Enc}^\prime$ is an encoding procedure for the code
conjugated to the Reed-Muller-code, i.e., for the code with the $X$-
and the $Z$-block of the stabilizer interchanged. This code is of CSS
type, like the Reed-Muller code itself. The encoding procedure changes
when passing through the Hadamard-gate because the encoded
Hadamard-gate is not local for the Reed-Muller quantum code. 

At any rate, the state  $|\overline{\Psi}_\text{algo}\rangle_S$ is
l.u. equivalent to a CSS-state encoded with CSS-codes, i.e., to a larger
CSS-state,  $|\overline{\Psi}_\text{algo}\rangle_S = \left(\bigotimes_{q \in
  {\cal{C}}_\text{odd} \cap Q} \bigotimes_{i \in S(q)} H_i\right)
|CSS^{\prime\prime\prime}\rangle_S$. Every CSS-state is
l.u. equivalent to a bi-colorable graph state \cite{Rains}, by a set
of local Hadamard-transformations. Thus, we finally obtain 
\begin{equation}
  |\overline{\Psi}_\text{algo}\rangle_S = \bigotimes_{i \in S_H}H_i
   \,\, |\Gamma\rangle_S.
\end{equation}
Therein, $S_H$ is some subset of $S$ and $|\Gamma\rangle_S$ is a
bi-colorable graph state with adjacency matrix of the corresponding graph
\begin{equation}
  \Gamma = \left(\begin{array}{cc} 0 & G^T\\ G & 0\end{array}\right).
\end{equation}

\paragraph{Circuit for a bi-colorable graph state.}
\label{BCGcirc}

\begin{figure}
  \begin{center}
    \epsfig{width=14cm, file=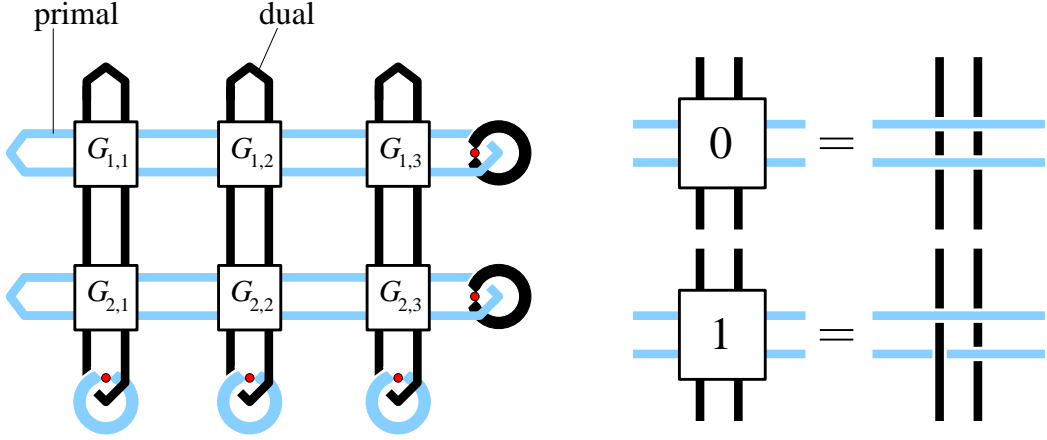}
    \caption{\label{circ}Creation of an arbitrary bi-colorable graph
    state. The singular qubits are displayed as red circles. $G$
    denotes the sub-matrix of the adjacency matrix $\Gamma$ which
    encodes the edges between primal and dual vertices.}
  \end{center}
\end{figure}

The circuit layout for the topologically protected creation of an arbitrary
bi-colorable graph state is shown in Fig.~\ref{circ}. The circuit
consists of a set of horizontal primal and a set vertical dual
defects. Each primal defect comes close to each dual defect 
once, and the two defects may be linked in that region. The form of
this junction is decided by a corresponding element of the graph state
adjacency matrix $G$. If $G_{i,j}=1$ then the defects are linked and
otherwise they are not. In addition, each of the loop defects
is linked with an ear-clip shaped defect which holds an $S$-qubit. The
graph state in question is formed among these qubits after the
remaining qubits have been measured.  

\begin{figure}
  \begin{center}
    \begin{tabular}{ll}
      a) & b)\\
      \parbox{8.5cm}{\epsfig{width=7cm, file=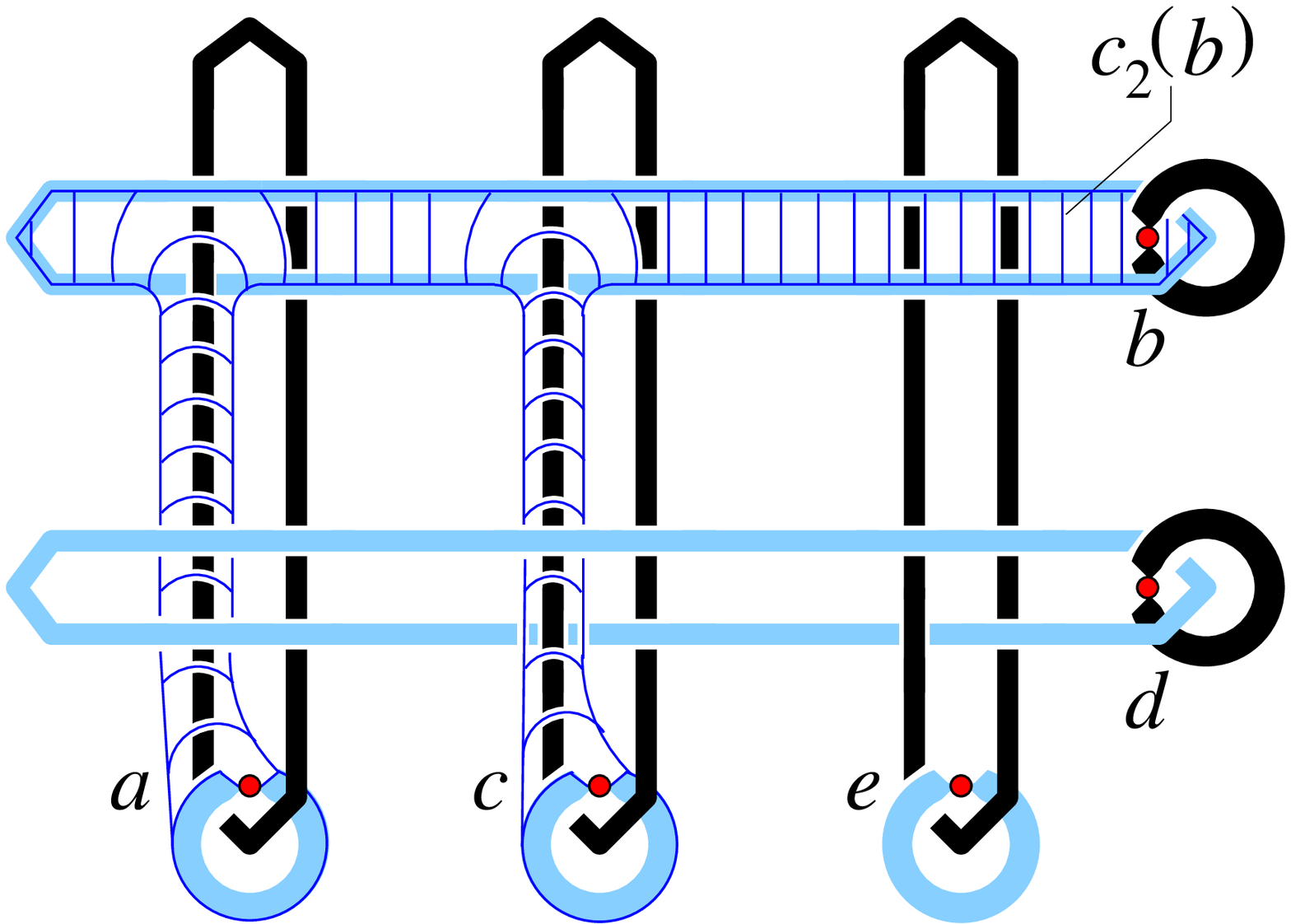}} &
      \parbox{5.2cm}{\epsfig{width=5.2cm, file=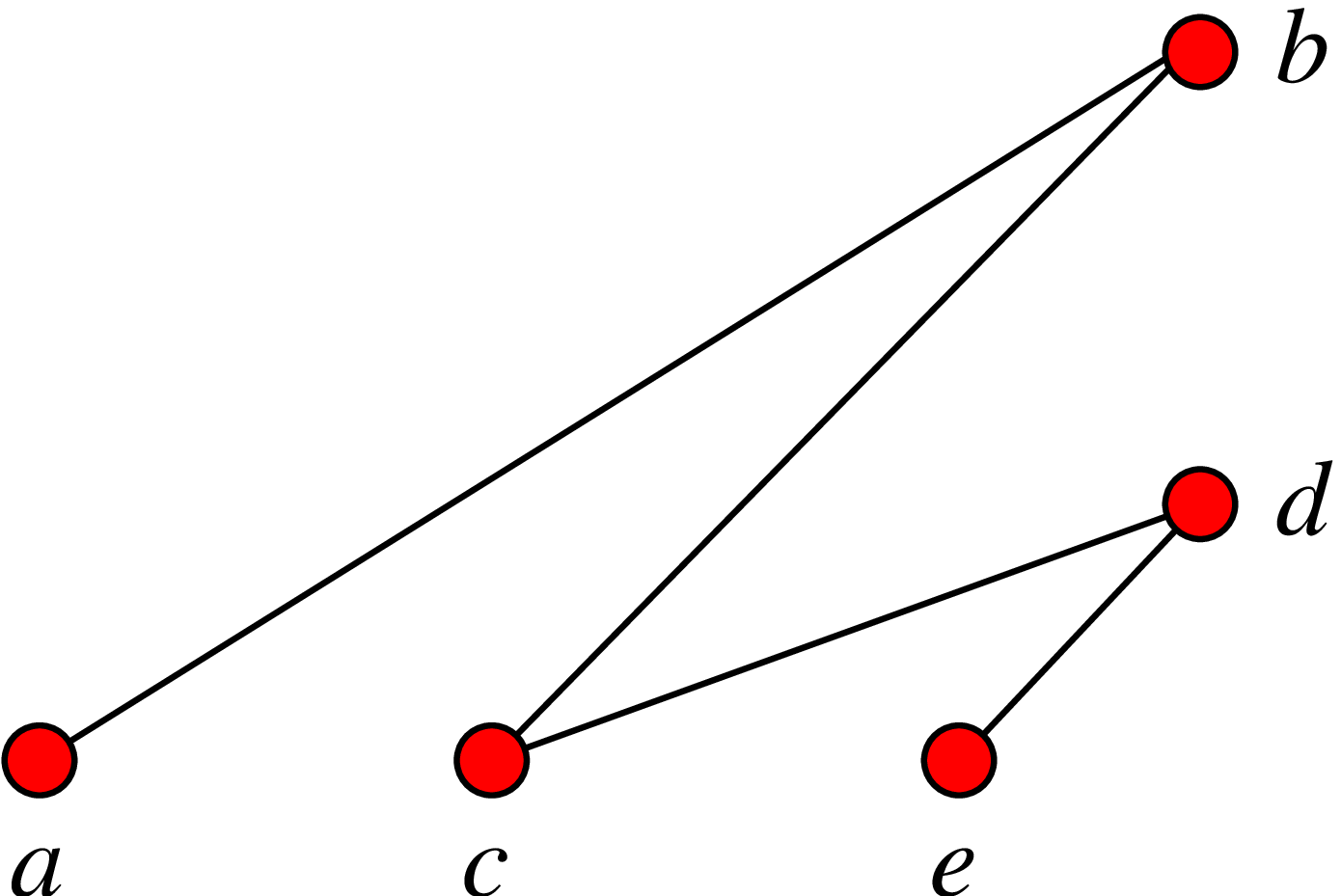}}
    \end{tabular}
    \caption{\label{circEx}Explanation of the circuit in
    Fig.~\ref{circ} for a particular example, the line graph. a)
    The relative 2-cycle $c_2(b)$ which gives rise to the
    graph-state stabilizer $K_b=Z_aX_bZ_c$. b) Graph corresponding to
    the created graph state.}
  \end{center}
\end{figure}

An explanation of the circuit in Fig.~\ref{circ}, for the specific
example of a line graph, is given in Fig.~\ref{circEx}. From this
example it should be clear how the circuit works in general. For the
line graph we have the adjacency sub-matrix \vspace{-3mm}  
\begin{equation}
  G_\text{line} = 
  \begin{array}{cl} \mbox{ }\\
    \left(\begin{array}{ccc} 1 & 1 & 0\\ 0 & 1 & 1
    \end{array} \right) & \begin{array}{c} b \\ d \end{array} \vspace{2mm}\\
    \begin{array}{ccc} a & c & e \end{array} \end{array}.
\end{equation}
This implies, for example, that in the circuit of Fig.~\ref{circEx}a
the dual defect winding around the 
dual $S$-qubit $a$ will be linked with the primal defect winding around
the primal $S$-qubit $b$. Now we explain how the
stabilizer element $K_b=Z_aX_bZ_c$ for the graph  state
$|\Gamma_\text{line}\rangle$ emerges. The other stabilizer generators
emerge in the same way.

Consider the relative 2-cycle $c_2(b)$ and imagine
it being built up step by step. We start around the $S$-qubit
$b$. Because $c_2(b)$ is primal and $b$ is primal, $K(c_2(b))$ affects
qubit $b$ by a Pauli-operator $X$, see Eq.~(\ref{KaffeS}). Also, $c_2(b)$
is bounded by the primal defect encircling $b$. 

We move further to the left. At some point, $c_2(b)$ approaches a dual
defect. Primal correlations cannot end in dual defects. Therefore,
$c_2(b)$ bulges out and forms a tube wrapping around the dual defect,
leading downwards. It ends in the primal defect holding the
dual $S$-qubit $c$. $K(c_2(b))$ affects qubit $c$ by a Pauli-operator
$Z$. Back at the junction, $c_2(b)$ continues to expand to the left. It
approaches a second dual defect where it forms another tube. In
result, the dual $S$-qubit $a$ is affected by a
Pauli-operator $Z$. Further to the left, the primal defect closes up and bounds
$K(c_2(b))$. 

$K(c_2(b))$ takes the form $X_b Z_a Z_c \bigotimes_{d \in V(b)V} X_d$
for some set $V(b) \subset V$. All qubits in $V$ are measured in the
$X$-basis (\ref{MP}) such that after these measurements the
correlation $K_b = \pm Z_aX_bZ_c$ remains. This is a stabilizer
generator for the graph state in Fig.~\ref{circEx}b
since $(a,b)$ and $(c,b)$ are the only edges of the line graph ending in
the vertex $b$. For every edge in the graph there is a link among
defect loops in the circuit.

The proof for the programmable circuit of Fig.~\ref{circ} realizing a general
bi-colorable graph state is similar. As an outline, each stabilizer
generator is associated with a 
doughnut-shaped defect in the circuit. Such a defect bounds a
correlation, and this correlation affects one $S$-qubit by
$X$. Further, because the considered defect is linked with other
defects of the opposite kind, the correlation surface forms
tubes. These tubes affect one other $S$-qubit each---the neighbors of the
first---by Pauli-operators $Z$. 

\paragraph{Implementing the local Hadamard-transformations.}

The equivalence between the graph state $|\Gamma\rangle_S$ and the
encoded algorithm specific resource
$|\overline{\Psi}_\text{algo}\rangle_S$ is by
Hadamard-transformations on some subset $S_H$ of $S$-qubits. Wherever such a
Hadamard-transformation needs to be applied, attach an extra loop to
the circuit in Fig.~\ref{circ},
\begin{equation}
  \parbox[c]{6cm}{\epsfig{width=5.8cm, file=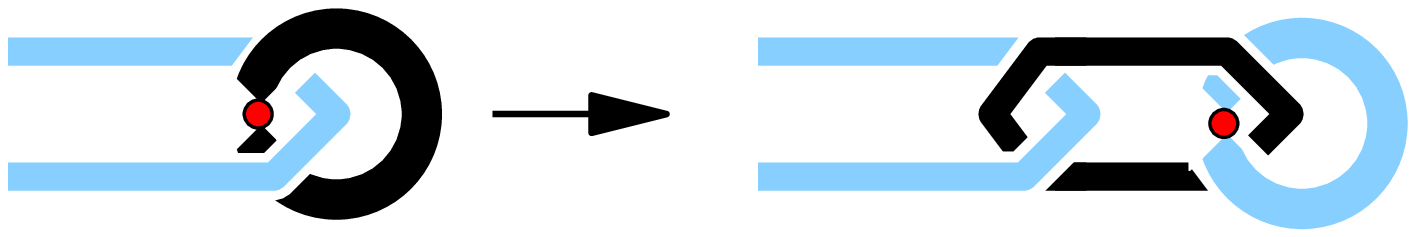}}.
\end{equation}

\section{Error sources, error correction and fault-tolerance threshold}

\subsection{Two error models}
\label{EM}

Below we describe two error models, a simple one and a more realistic one. 

\begin{Model}The cluster state is created only on those cluster qubits
  which are needed for the computation, i.e., on $V \cup S$. The defect
  qubits of $D$ are left out.
\begin{enumerate}
        \item{The noise is described by independent partially
  depolarizing channels acting on each cluster qubit. The noisy state
  $\rho_{\cal{C}}$ is given 
  by $\displaystyle{\rho_{\cal{C}}=\bigotimes_{a \in
  {\cal{C}}}T_a (p_1)\, 
            |\phi\rangle_{\cal{C}}\langle\phi|}$, with
	\begin{equation}
	  \label{LDC}
	  T_a^{(1)}(p_1)=
	  (1-p_1)[I_a]+ \frac{p_1}{3}\left([X_a]+[Y_a]+[Z_a]\right).
	\end{equation}}        
        \item{The classical computation for syndrome processing is
        instantaneous.}
\end{enumerate}
\end{Model} 
The reason for considering this error model first is its simplicity. We
would like to separate the intricacies inherent in the presented
error-correction scheme from
additional difficulties incurred by a realistic error model. The basic
justification for such an approach is this: The error correction used
here is topological. Therefore, a threshold should exist regardless of
whether independent errors are strictly local or only local in the sense of
having a support of bounded size.    

The most straightforward method to create a cluster state is
from a product state $\bigotimes_{a} |+\rangle_a$ via a
(constant depth) sequence of $\Lambda(Z)$-gates \cite{BR}. If these
gates are erroneous, then Error Model 1 does not apply 
in general.\footnote{There may be situations in which the
error Model 1 is in fact a 
good approximation. For example, consider a scenario in which the
cluster state is purified before being measured for computation. Of
course, the gates 
in a purification protocol would be erroneous, too, such that the
purified state is not perfect. In effect, the errors of the initial
state were replaced by the errors of the purification protocol. There
exist purification protocols \cite{HA} in which the gates act
transversally on two copies of the cluster state (one 
of which is subsequently measured). As a result, the errors introduced 
by the purification are approximately local, as in Error Model 1.  The
purification 
protocol \cite{HA} in its current form has a problem of its own, though; due
to the exponentially decreasing efficiency of post-selection, it is not
scalable in the size of the state. But chances are that this can be
repaired.} Specifically, one may raise the following objections to
Error Model 1:
\begin{itemize}
\item{No correlated  errors are included in Error Model 1. Creating
  the cluster state via a sequence of gates will, however, lead
  to correlated errors in the output cluster state.}
\item{Storage errors accumulate in time. There is temporal
  order among the measurements such that 
  the computation takes a certain time $t_\text{comp}$ which cannot be
  bounded by a constant for all possible computations. As
  a consequence, for
  the qubits measured in the final round the local noise rate
  increases monotonically with $t_\text{comp}$ and exceeds
  the error threshold.}
\item{To leave the $D$-qubits out is a deviation from the
  originally envisioned setting: the cluster state
  on  $S  
  \cup V$ is algorithm-specific.\footnote{For the creation of the cluster
  state this makes little or no difference: fewer
  gate operations are needed than for the creation of
  $|\phi\rangle_{\cal{C}}$. For parallelized procedures that make use of the
  translation invariance of ${\cal{C}}$, it should not be too difficult
  to remove the superfluous $D$-qubits from the lattice before
  the remaining qubits are entangled.}}    
\end{itemize} 
To account for these inadequacies, we consider a
second error model.

\begin{figure}
  \begin{center}
    \epsfig{width=9cm, file=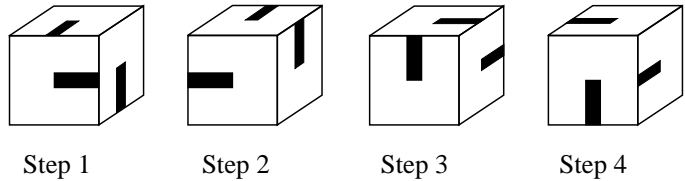}
    \caption{\label{GateSeq}Steps of $\Lambda(Z)$-gates for the
    creation of a cluster state on a bcc-symmetric lattice, for the
    sub-clusters ${\cal{C}}_k$. For $k$ odd, the sequence is $1
    \rightarrow 2 \rightarrow 3 \rightarrow 4$, and for $k$ even it is
    $3 \rightarrow 4 \rightarrow 1 \rightarrow 2$.} 
  \end{center}
\end{figure}

\begin{Model}
  A cluster state on a bcc-symmetric lattice is
  created in four steps 
  of nearest-neighbor $\Lambda(Z)$-gates. The gate sequence is
  as shown in Fig.~\ref{GateSeq}. Errors
  occur due to the erroneous preparation of the initial
  $|+\rangle$-qubits, erroneous $\Lambda(Z)$-gates in the process of
  creating the cluster state, storage and measurement.
  \begin{enumerate}
  \item{The computation is split up into steps which performed on
    sub-clusters ${\cal{C}}_k$. In each step,
    unmeasured qubits remaining from the previous step---the hand-over
    qubits---are loaded into a cluster state on a sub-cluster.
    Subsequently, all but a few cluster qubits (the new
    hand-over qubits) are measured. The steps have their temporal depth 
    adjusted such that each qubit, after being locally prepared and
    entangled, waits at most a constant number $t_0$ of
    time steps until its measurement occurs, $t_0\geq1$. Error in
    storage is described by a partially depolarizing 
    noise with error probability $p_S$ per time step.}
  \item{The erroneous preparation of initial $|+\rangle$-qubits is
    modeled by the perfect procedure followed by local depolarizing
    noise {\em{(\ref{LDC})}}, with
    probability $p_P$.  Measurement is described by perfect
    measurement preceded 
    by partially depolarizing noise with error 
    probability $p_M$. The erroneous $\Lambda(Z)$-gates are modeled
    by the perfect gate followed by a 2-qubit depolarizing channel
    \begin{equation}
      \label{T2}
      T^{(2)}_{e,f}(p_2) = (1-p_2) [I_{e,f}] + \frac{p_2}{15}\left([I_e \otimes
      X_f] + ... + [Z_e \otimes Z_f] \right).
     \end{equation}} 
  \item{Classical syndrome processing is instantaneous.}
  \end{enumerate}
\end{Model}
In the subsequent sections we compute a fault-tolerance threshold
for both error models. In Error Model 2, storage error will cause the
minimum damage for the smallest possible value of $t_0$, which is $t_0=1$. 
In this case, each sub-cluster carries a subset of $S$-qubits with no
mutual temporal dependence. All qubits in $V$ and $D$, except the
hand-over qubits (See Appendix~\ref{SubClusters}), are measured
immediately after being entangled. The measurement of the $S$-qubits
has to wait one time step. {\em{Henceforth we set $t_0=1$.}}

\subsection{Methods for error correction and the threshold value}
\label{MECTV}

There are different methods of error correction associated with the 
different regions $V$, $S$ and $D$ of the cluster. In $D$, there are 
as many inequivalent errors as there are syndrome bits, such that the error
correction is trivial. The error correction in $V$ is based 
on the random plaquette $\mathbb{Z}_2$-gauge model in three 
dimensions \cite{RPGM}. The error correction in $S$ is carried out 
using the (concatenated) quantum Reed-Muller code. 

\paragraph{Error correction in $D$.}In the domain $D(d)$ of the defect only
$Z$-errors matter for face qubits and only
$X$-errors matter for edge qubits. Any other errors may be absorbed into
the subsequent 
measurements (\ref{MP}). The $X$-errors on edges may be relocated to
$Z$ errors on the 
neighboring face qubits via $X_e \cong K_e X_e$, such that we
need to consider $Z$-errors on face qubits only. For each face in $d$ we learn
one syndrome bit, yielding a unique syndrome for each error configuration. 

However, if an error $X_e$ on an edge qubit $e$ in the surface of defect
$d$ is relocated, the equivalent error $K_e X_e$ may partially be outside
$D(d)$. After error correction in $D(d)$, an individual
$Z$-error on a face qubit in $V$ is left next to $D(d)$. This affects
the error correction in $V$ near $D$; see below.

\paragraph{Error correction in $V$.}First consider a scenario where
the entire cluster consists of the region $V$ ({\it i.e.}, there are no
defects and no singular qubits). Error correction on the primal
lattice ${\cal{L}}$ and the dual lattice $\overline{\cal{L}}$ run
separately. Here we consider error correction on the primal
lattice only; error correction on the dual lattice is analogous.

The error chains live on the edges of the lattice ${\cal{L}}$ and
leave a syndrome at the end points, which are vertices of
${\cal{L}}$. This is exactly the scenario which has been considered
for topological quantum memory in \cite{RPGM}, and subsequently
the results of \cite{RPGM} that we
need in the present context are briefly summarized. The connection
between topological error 
correction and cluster states has been made in \cite{LRE} for the
purpose of creating long-range entanglement in the presence of noise. 

Given a particular syndrome and an error chain $E(c_1)$
compatible with this syndrome, we are interested in the total
probability $P(c_1)$ of the {\em{homology class}} of $c_1$,
\begin{equation}
  P(c_1) = \sum_{z_1 \in Z_1} p(c_1+z_1), 
\end{equation}
where $p(c_1^\prime)$ is the probability of an individual error chain
$E(c_1^\prime)$, and the sum is over all 1-cycles. For error correction we
infer that the physical error which occurred was from the homology
class with the largest probability.

If the errors on the lattice edges occur
independently with a probability $q$ then the problem of computing
$P(c_1)$ for a given chain $c_1$  can be mapped onto a problem from
statistical mechanics, namely the random plaquette $\mathbb{Z}_2$-gauge 
model in three dimensions \cite{RPGM}. The crossover from 
high fidelity error correction at small error rates to low fidelity
error correction at high error rates corresponds to a phase
transition in this model. A numerical estimate of the critical error rate is
$q_{\text{c}}=0.033\pm0.001\;$\cite{Ohno}. 

As far as is known, the classical operational resources required to
find the most likely error homology class consistent with a given
syndrome scales exponentially in the number of error locations. The
assumption of the classical processing being instantaneous cannot be
justified under these conditions. However, it is possible to trade
threshold value for efficiency in the error correction procedure. A
reasonable approximation to the maximum probability for a
homology class of errors is the probability of the lowest 
weight admissible chain. The minimum-weight perfect matching algorithm
\cite{MWCM,CR} computes this chain using only
polynomial operational resources. A numerical estimate to the
threshold with this algorithm for error correction is
$q^\prime_{\text{c}}=0.0293\pm0.0002$ \cite{WHP}.  

{\em{Remark:}} The topological error threshold is estimated in numerical
simulations of finite-size systems. 
For this purpose, the probability of logical error is plotted vs. the physical
error parameter for various system sizes. 
For sufficiently large lattices (such that finite-size effects are small),
we expect these curves to follow a universal scaling ansatz near the threshold
such that they share a common intersection point and their slopes
are proportional to a common power of the lattice size.
As the system size is increased to infinity, we then expect the curves
to approach a step function which transitions at the threshold 
value of the physical error parameter.\medskip  

The above quoted threshold value is for independent errors on the
edges of ${\cal{L}}$. Do the models for the physical error sources  
of Section~\ref{EM} lead to such independent errors? The
answer is `yes' 
for error Model 1 and `no' for error Model 2. For the latter, we 
need to consider a modified RPGM with correlated errors among
next-to-nearest neighbors. Specifically, for Error Model 1
the relation between the local rate $p_1$ of the physical depolarizing
error and the error parameter $q$ that shows up in the RPGM is
\begin{equation}
  q=\frac{2}{3}p_1, \;\;\;\;\;\;\;\;\text{for Error Model 1}.
\end{equation}
Given a threshold of $q = 2.93\%$ \cite{WHP} for error correction 
via the minimum weight perfect matching algorithm in the bulk, 
then the corresponding depolarizing error rate that can be tolerated is 
\begin{equation}
  \label{Trpgm_M1}
  p_{1,\text{c}} = 4.4\times 10^{-2}, \;\;\;\; \mbox{(in $V$)}.
\end{equation} 

\begin{figure}
  \begin{center}  
    \begin{picture}(9,8)(0,0)
    \put(-1.2,8){\epsfig{width=8cm, angle=270, file=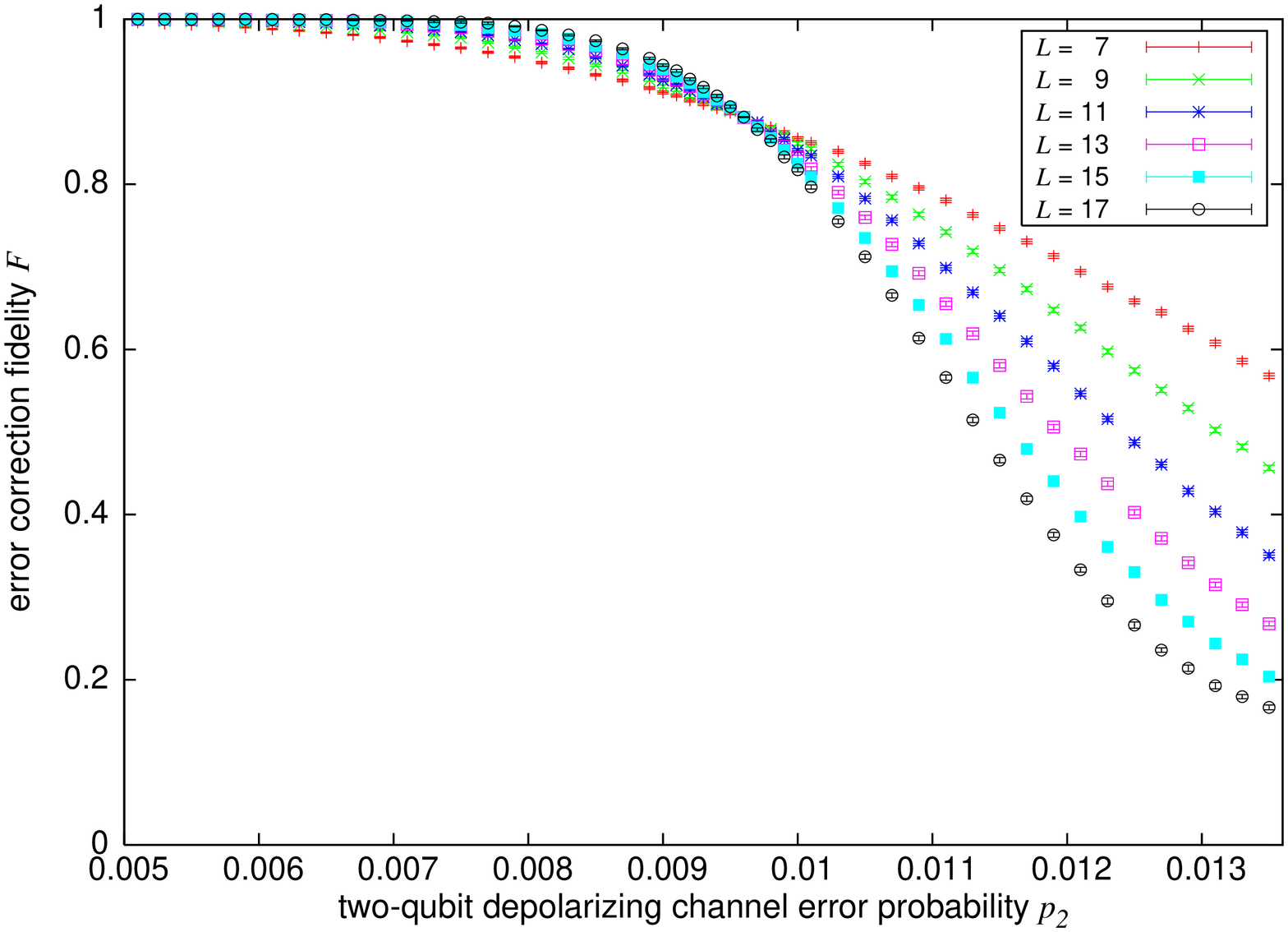}}
    \put(0.4,5.3){\epsfig{width=4cm, angle=270, file=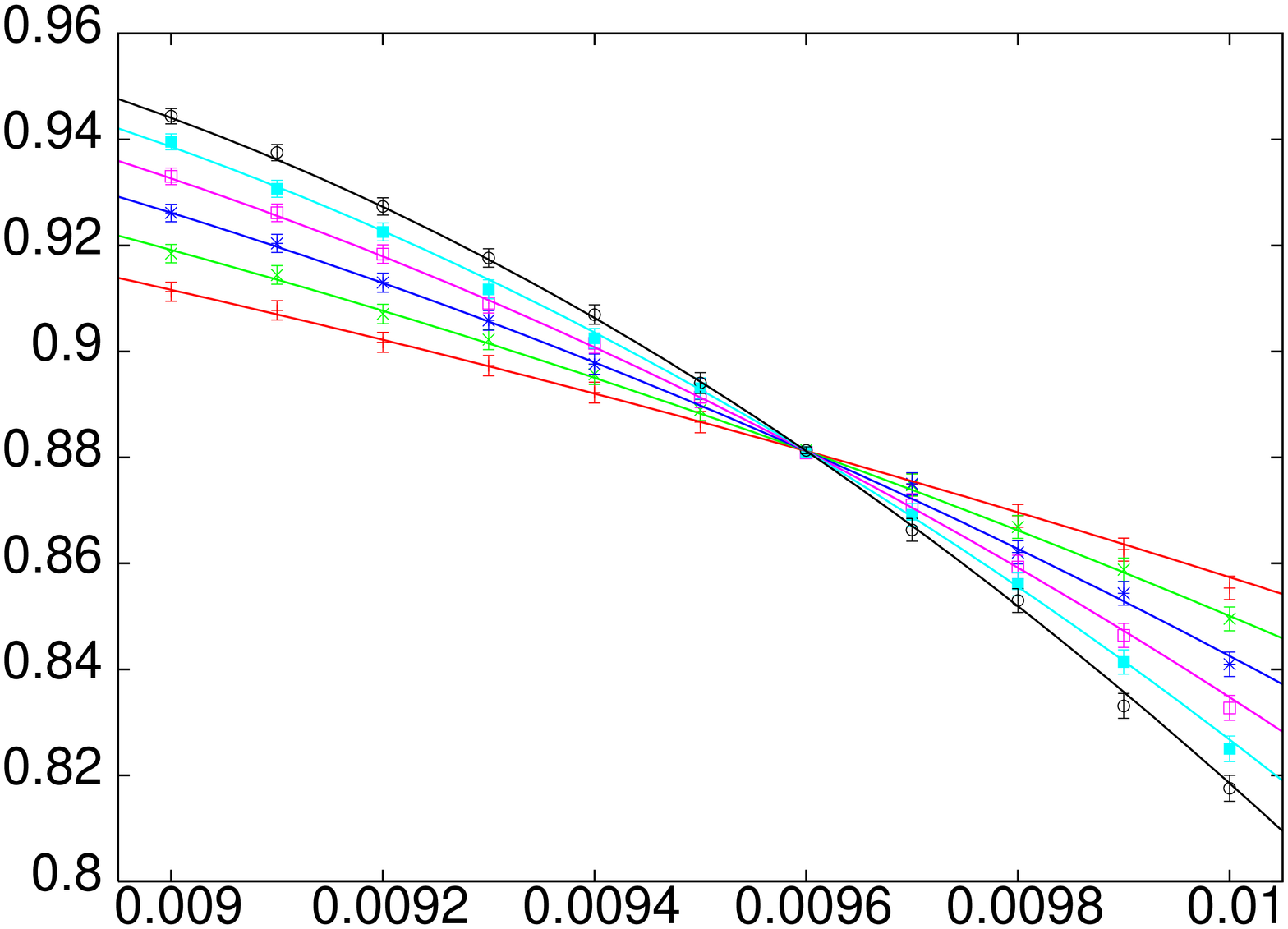}}
    \end{picture}
    \caption{\label{TE}Threshold estimation in lattices of finite
    size, for periodic boundary conditions. Here, with
    $\Lambda(Z)$-gates as the only error source,     
    we find a threshold of $p_{2,\text{c}} = 9.6\times 10^{-3}$.
    In the inset, best fits to the universal scaling ansatz are drawn.  
    Error bars denote two standard deviations due to finite sampling
    size. } 
  \end{center}
\end{figure}

For Error Model 2, first consider the case where only the
$\Lambda(Z)$-gates are erroneous,
$p_P=p_S=p_M=0$. Then, in addition to local errors with a
rate $q_1$ there exist correlated errors with error rate $q_2$ for
each pair of opposite edges in all faces of ${\cal{L}}$. That is, with
a probability  $q_2$ simultaneous errors are introduced on opposite
edges of the faces in ${\cal{L}}$.  The local noise specified
by $q_1$ and the two-local noise specified by $q_2$ are independent
processes.  The relations between the error parameter $p_2$ of 
the $\Lambda(Z)$-gates and the parameters $q_1, q_2$ of the RPGM with
correlated errors are
\begin{equation}
  \label{RPGMmap}
  \begin{array}{rcl}
    q_1 &=& \displaystyle{\frac{32}{15}p_2 \left( 1-\frac{8}{15}p_2 \right)
    \left(\frac{64}{225}p_2^{\,\,2} + \left(1-\frac{8}{15}p_2\right) ^2
    \right),}\\
    q_2 &=& \displaystyle{\frac{1}{2}-\sqrt{\frac{1}{4}-\frac{4}{15}p_2}\, =\,
    \frac{4}{15}p_2 + O(p_2^{\,\,2}).}
  \end{array}
\end{equation} 
The correlation of errors on sites separated by a distance
of two arises through error propagation in the creation of the cluster
state. Correlations among errors on next-neighboring sites play no
role because such errors  live
on different lattices (${\cal{L}}$ and $\overline{\cal{L}}$) and  are
corrected independently.

The only effect of $p_P,p_S,p_M>0$ is an enhanced local error
rate $q_1$. We give the relations to leading order only; they read
\begin{equation}
  \label{RPGMmap2}
  \begin{array}{rcl}
    q_1 &=&
    \displaystyle{\frac{32}{15}p_2+\frac{2}{3}\left(p_P+p_S+p_M\right),}
    \vspace{1mm}\\  
    q_2 &=& \displaystyle{\frac{4}{15}p_2 ,}
  \end{array}\;\;\;\;\;\;\;\;\;\;\; \text{for error Model 2}.
\end{equation}
See Fig.~\ref{TE} for a simulation of error correction under faulty
$\Lambda(Z)$-gates 
as the only error source, which gives rise to correlated noise between
neighboring edge qubits.  If we define $x = (p_2 - p_{2,c}) L^{1/\nu_0}$
then the universal scaling ansatz states that fidelity $F$ should be a function
dependent only on the scaling parameter $x$ in the vicinity of the
threshold \cite{WHP}. 
We find very good agreement (with $R^2 > .9991$) for $F = A + Bx + Cx^2$,
where we fit for constants $A$, $B$, $C$, $p_{2,c}$, and $\nu_0$.  This gives
very tight bounds on the critical probability $p_{2,c} = 9.6 \times 10^{-3}$.
Interestingly, we also find $\nu_0 = 1.00 \pm 0.02$, which indicates that 
this model belongs to the same universality class as the purely local
error model 
of the 3D-RPGM \cite{WHP}.

\begin{figure}
  \begin{center}
      \epsfig{width=7.5cm, angle=270,
	file=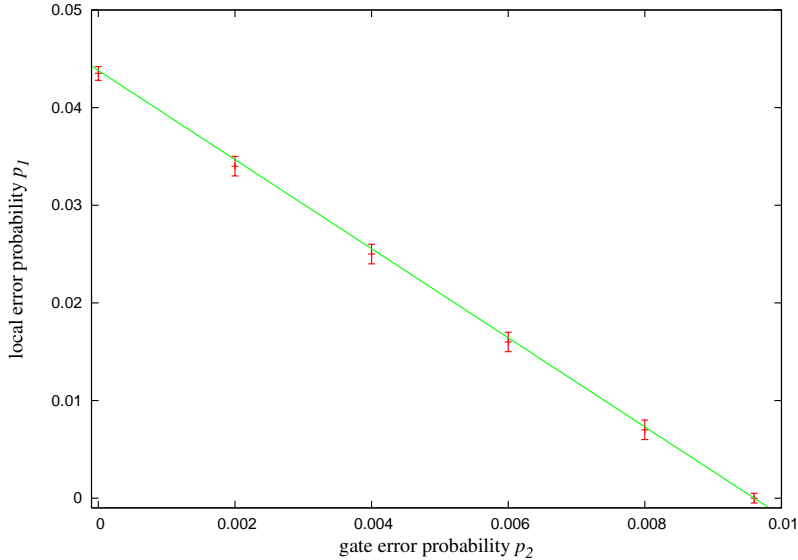}  
    \caption{\label{NT}Trade-off curve for the threshold
    value in the presence of local and two-local errors. Horizontal
    axis: two-qubit gate error $p_2$, vertical axis: 
    local error rate $p_P+p_S+p_M$.} 
  \end{center}
\end{figure}

In Fig.~\ref{NT} the threshold trade-off curve between $p_P+p_S+p_M$
and $p_2$ is displayed. Numerically, we obtain for the thresholds in the bulk
\begin{equation}
  \label{Trpgm_M2}
  \begin{array}{rclcl}
    p_P=p_S=p_M&=& 1.46 \times 10^{-2}, && \mbox{for }p_2=0,\\
    p_2&=&0.96 \times 10^{-2}, && \mbox{for }p_P=p_S=p_M=0,\\
    p_i&=&0.58\times 10^{-2},  && \mbox{for }p_P=p_S=p_M=p_2.
  \end{array}
\end{equation}

\paragraph{Error correction in $V$ near $D$.}
In the presence of defects there are two modifications to error
processes in $V$.  
First, the length scale for the minimum extension of a non-trivial error cycle
shrinks. Second, there is a surface effect; the effective error rate
for qubits in $V$ next to the surface of defects is enhanced by a
constant factor.

1. Length scale for non-trivial errors: For comparison, consider a
cluster cube of 
finite size $2L\times 2L \times 2L$. A non-trivial error cycle
must stretch across the entire cube and thus has a weight of at least
$L$. The lowest weight errors which are misinterpreted by the error
correction procedure occur with a probability $q^{L/2}$. The total
error probability incurred by such errors may therefore be expected to
decrease exponentially fast in $L$, which is confirmed in numerical
simulations \cite{LRE}.  

\begin{figure}
  \begin{center}
    \epsfig{width=5.5cm, file=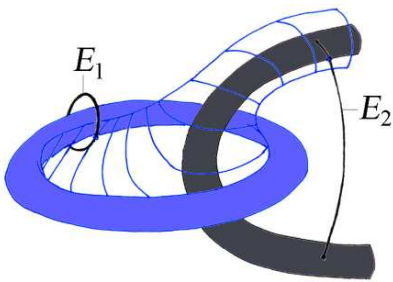}
    \caption{\label{ES}Sources for logical error in the presence of
    defects. } 
  \end{center}
\end{figure}

In the presence of defects, the dominant sources for logical error are
1-chains that either wind around a defect or that begin and end in
a defect and intersect a correlation surface (2-chain) in
between; see Fig.~\ref{ES}. The relevant length scales are thus the
thickness (circumference) and the 
diameter of the defects. They are much smaller than $L$.

Specifically, consider a defect with circumference $u$ and length $l$
which bounds a correlation surface $c_2$, such that $|\{\partial c_2\}|=l$. An
error cycle winding around 
the defect has a weight of at least $u+8 \approx u$, and there are $l$ such
minimum weight cycles. Therefore, the probability $p_E(u,l)$ for affecting
$K(c_2)$ by an error is, to lowest contributing order,
\begin{equation}
  \label{expsupp}
  p_E(u,l)= l\frac{u!}{(u/2)!^2} q^{u/2} \approx l \exp\left(\frac{\ln
  4q}{2} u \right) \frac{1}{\sqrt{\pi/2\, u}}. 
\end{equation}
In the range of validity for the above expansion in powers of $q$, the
error is still exponentially suppressed in the relevant length scale $u$.   

2. Surface effects: As discussed above, the error level is enhanced
for qubits in $V$ near the 
surface of a defect. If the defect is primal (dual), the
enhancement occurs on dual (primal) qubits. This effect will---if
anything---lower the threshold. But there is another effect: the presence
of the defect changes the boundary conditions. In case of a primal defect,
the boundary conditions  on the defect surface become rough for
the primal lattice and smooth for the dual lattice. Dual error chains
cannot end in a primal defect, as we noted earlier. For the dual
lattice, there is excess
syndrome available at the defect surface. This effect will---if
anything---increase the threshold. Our intuition is that neither
effect has an impact on the threshold value. 
The threshold should, if the perturbations at the boundary are not
too strong, still be set by the bulk. 

We have performed numerical
simulations for lattices of size $L \times L \times 2 L$, where half
of the lattice belongs to $V$ and the other half to the defect
region $D$. The error rate is doubled near the mutual boundary
of the regions and there is no remaining error in $D$. Simulations are
feasible with reasonable effort  up to $L \approx 20$. 
We find that finite-size effects (due to the smooth boundary conditions)
are still noticeable up to these lattice sizes, but the intersection point of 
fidelity curves for nearby lattice sizes is slowly converging to a threshold 
value around that of the bulk ($\sim 2.9\%$).

\begin{figure}
  \begin{center}
    \begin{tabular}{lcl}
      a) & \mbox{ }& b)\\ \\
      \parbox[c]{8.4cm}{\epsfig{width=8.4cm, file=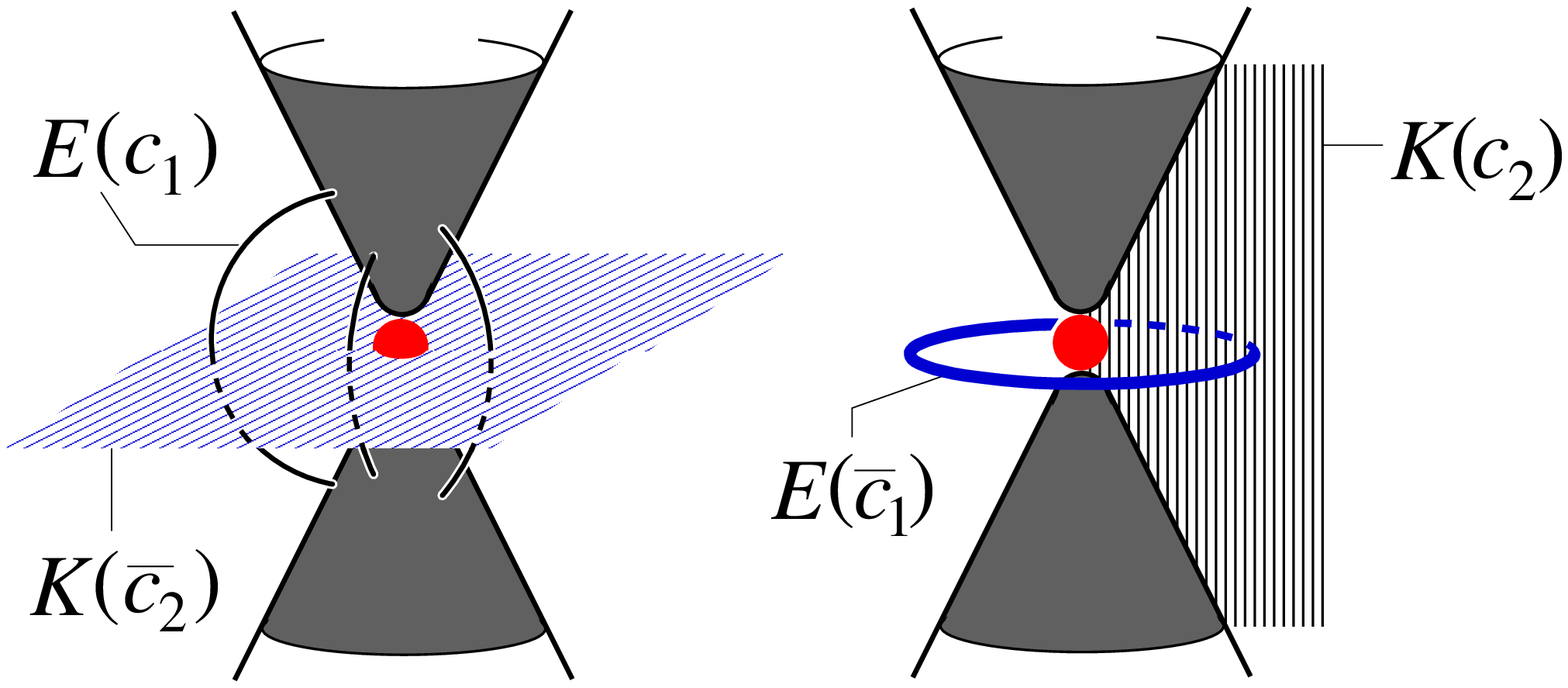}} && 
      \parbox[c]{5.6cm}{\epsfig{width=5.4cm, file=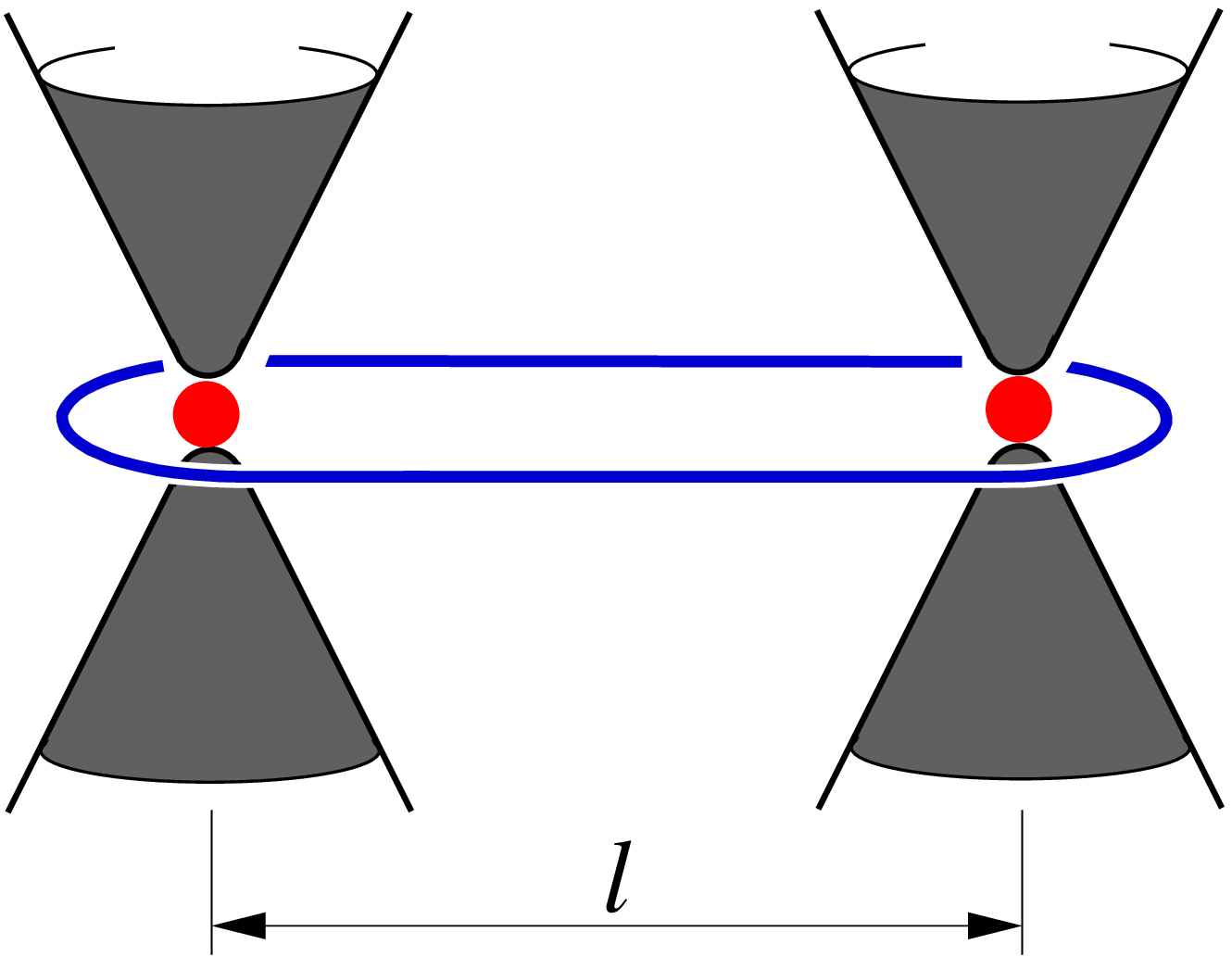}}
    \end{tabular}
    \caption{\label{NearS}Topological error correction in the presence
    of defects. a) Short error cycles intersect with the
    correlation surfaces and thereby jeopardize the
    topological error correction. b) Correlated errors on $S$-qubits
    are suppressed exponentially in the qubit separation $l$.}
  \end{center}
\end{figure}

\paragraph{Error correction in $V$ near a junction between $D$ and $S$.}
Near an $S$-qubit there exist relative error cycles of small length, see
Fig.~\ref{NearS}a, and the topological error correction breaks down. 
As a result, the effective error on an $S$-qubit is
enhanced by its surrounding. To compute the effective error
probabilities, we replace every low-weight error-chain $E(\gamma)$ that
results in a logical error {\em{after error 
correction}} by an equivalent error $E_S(\gamma)$ acting on the
$S$-qubits. The error correction converts $E(\gamma)$ into
$E(c_1(\gamma))$ with $c_1(\gamma)$ a relative 1-cycle. `Equivalent' means
that  $E(c_1(\gamma))$ and $E_S(\gamma)$ act in the same way on the
stabilizer generators $\{K_{\overline{\Psi},s}|\; s \in S\}$ of the
induced state $|\overline{\Psi}_\text{algo}\rangle_S$, i.e.,
$[E(c_1(\gamma))E_S(\gamma),K_{\overline{\Psi},s}]=0$ for all $s\in
S$. The relevant correlations to check are $K(\overline{c}_2)$ and
$K(c_2)$ displayed in Fig.~\ref{NearS}a.

It is important to note that the
effective error on the $S$-qubits is {\em{local}}. This arises because
only error chains causing a 1-qubit error may have small length. Error
chains causing a correlated error on the $S$-qubits are suppressed 
exponentially in the qubit separation. See Fig.~\ref{NearS}b.

\begin{figure}
  \begin{center}
    \begin{tabular}{ll}
      a) & b)\\
      \parbox{5cm}{\epsfig{width=5cm,file=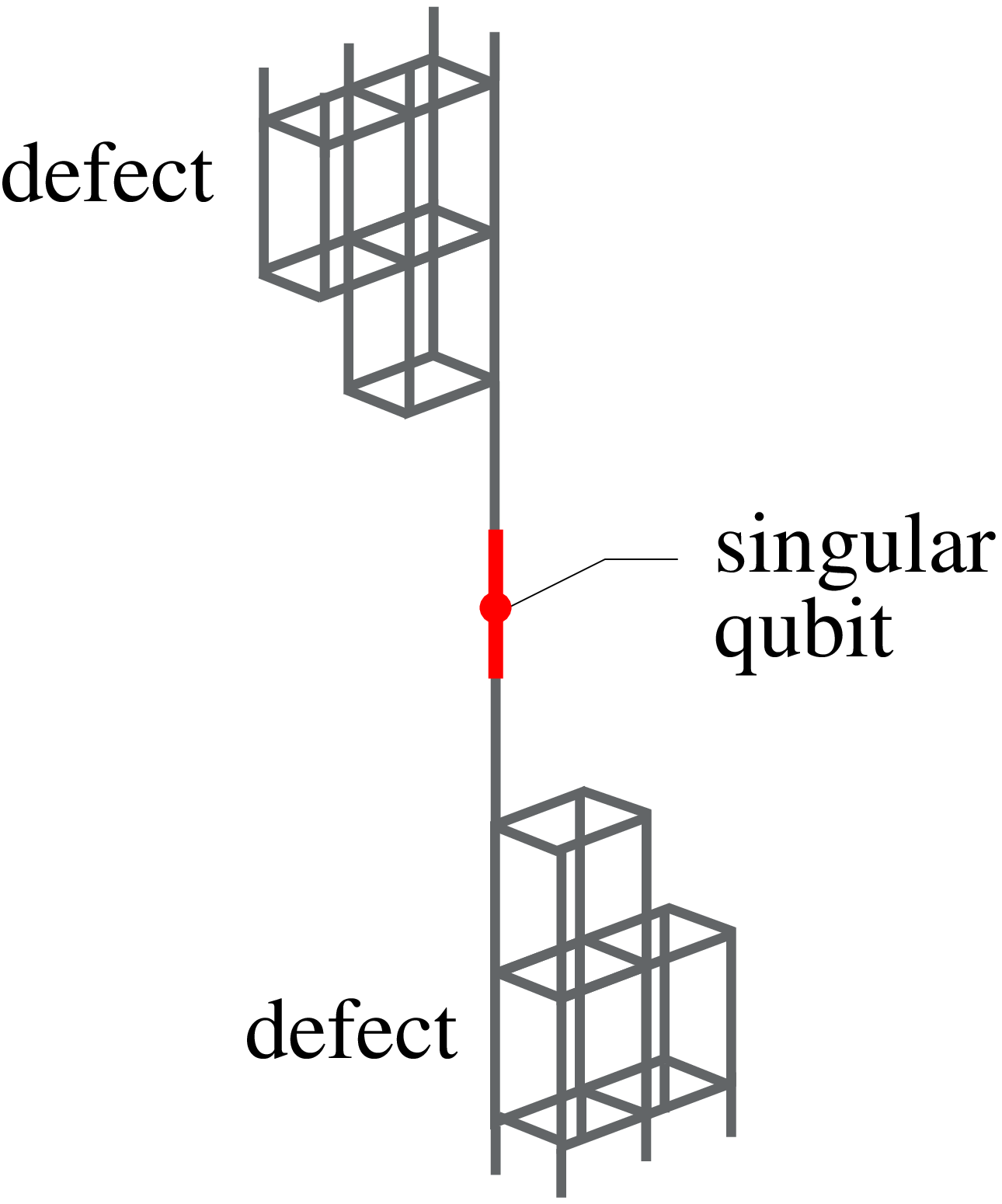}}&
      \parbox{6cm}{\epsfig{width=6cm, file=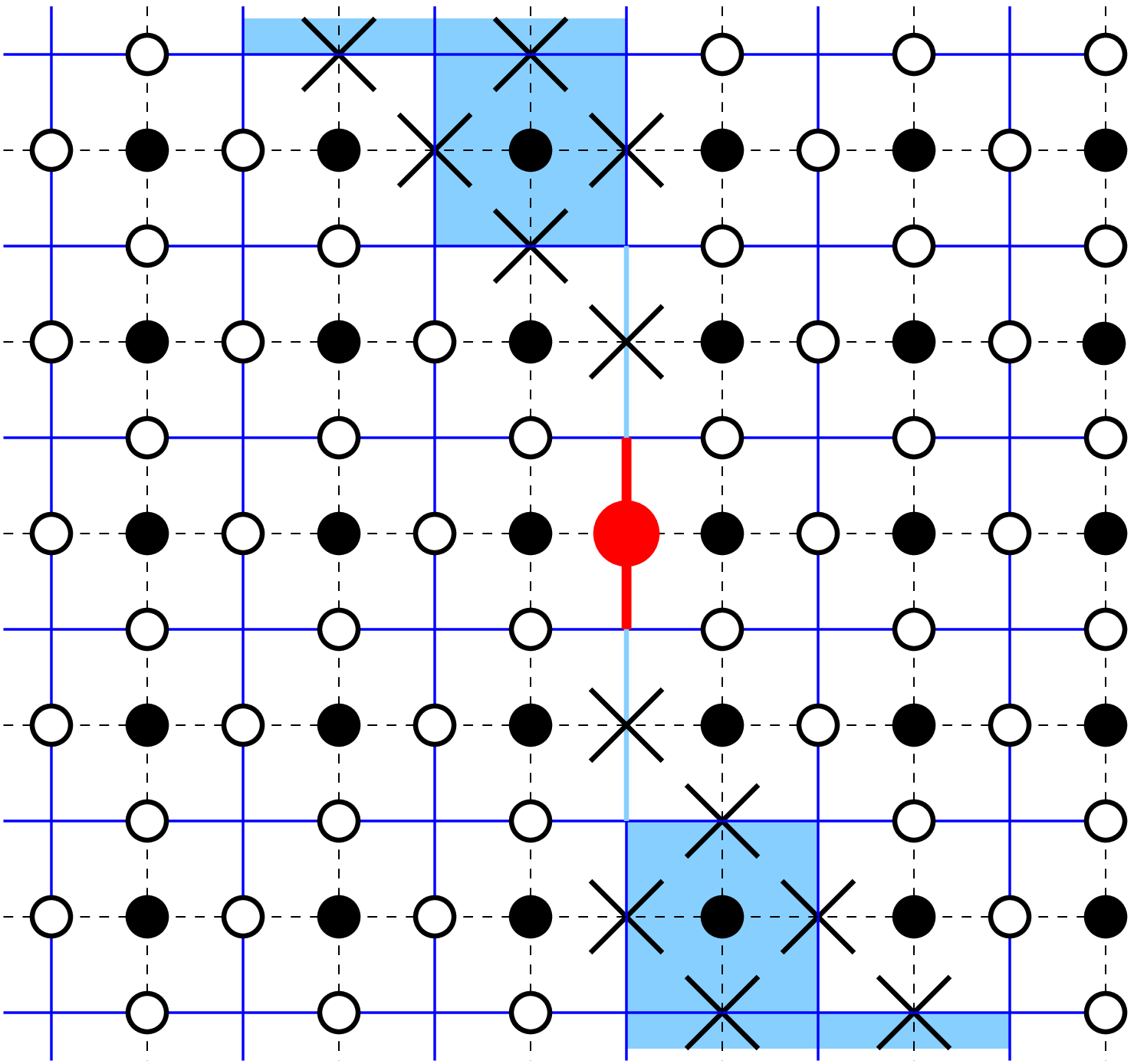}}
    \end{tabular}
    \caption{\label{Detail}Shape of the
    defect near an $S$-qubit (red). a) Three-dimensional view. b)
    Cross-section through the cluster.  Edges which
    belong to the defect are marked as ``$\times$''. Black dashed lines
    connect neighbors on the cluster. Blue underlay: faces in the
    defect. Even/odd qubits: $\bullet/\circ$. } 
  \end{center}
\end{figure}

We compute the effective error channel on
the $S$-qubit to first order 
in the error probabilities only. For error Model 1 the error
enhancement only affects sub-leading orders of $p_1$,
\begin{equation}
  \label{effSEC1}
    T^{(1)}_s =
    (1-p_1)[I_s]+\frac{p_1}{3}\left([X_s]+[Y_s]+[Z_s]\right).
\end{equation}
For Error Model 2 the effective error channel on an $S$-qubit is not
universal but depends on the precise shape of the defect double-tip near
the $S$-qubit. For our 
calculation we use the defect shape displayed in
Fig.~\ref{Detail}. The defect is one-dimensional nearest to the
$S$-qubit, and farther away becomes three-dimensional.  The effective
error channel then is  
\begin{equation}
\label{effSEC}
\begin{array}{rcl}
  \tilde{T}^{(1)}_s &=&
    \displaystyle{\left(1-\frac{2}{3}p_P-\frac{7}{3}p_S-\frac{7}{3}p_M
      -\frac{94}{15}p_2\right)[I_s]\,
    +\left(2p_2 + \frac{5}{3}p_S+\frac{5}{3}p_M\right) [X_s]\,+}\\
    &&
    \displaystyle{+\left(\frac{2}{5}p_2+ \frac{1}{3}p_S+\frac{1}{3}p_M \right)
    \,[Y_s]+\left(\frac{2}{3}p_P+\frac{1}{3}p_S+\frac{1}{3}p_M 
    +\frac{58}{15} p_2 \right) [Z_s]}. 
  \end{array}
\end{equation}
The individual contributions to  (\ref{effSEC}) are listed in 
Appendix~\ref{E2}. 

\paragraph{Error correction in $S$.}The $S$-qubits are protected by
the concatenated Reed-Muller code. This code 
corrects for the errors (\ref{effSEC1})/(\ref{effSEC})
that remain after error correction in $V$ and $D$.   

The $S$-qubits are all measured in the eigenbases of $\frac{X\pm
  Y}{\sqrt{2}}$. Then, an $X$- or a $Y$-error is equivalent to a
$Z$-error with half 
the probability. This is easily verified for the case where $X$- and
$Y$-errors occur with the same 
probability. W.l.o.g. assume the measurement basis is
$\frac{X+Y}{\sqrt{2}}$. Then 
$[X]+[Y]= \frac{[X+Y]}{\sqrt{2}} + \frac{[X-Y]}{\sqrt{2}} =
\frac{[X+Y]}{\sqrt{2}} ([I]+[Z]) \cong [Z]$. But the statement is, to
  leading and next-to-leading order in the error probability, also 
true when $X$- and $Y$-errors do not occur with equal probability; see
Appendix~\ref{E3}.  We may thus convert the $X$- and $Y$-errors in
(\ref{effSEC1}) and (\ref{effSEC}) into $Z$-errors. The corresponding
error probability $p_Z$ is
\begin{equation}
  \label{eqZerr}
  \begin{array}{rclcl}
    p_Z&=&\displaystyle{\frac{2}{3}{p_1}}, && \mbox{for error Model
    1},\vspace{1mm}\\ 
    p_Z&=&
    \displaystyle{\frac{76}{15}p_2+\frac{2}{3}p_P+\frac{4}{3}p_M+
      \frac{4}{3}p_S,} && \mbox{for error Model 2}.
  \end{array}
\end{equation}
The fault-tolerance threshold of the concatenated
$[15,1,3]$-Reed-Muller code for independent $Z$-errors with
probability $p_Z$ is $1.09 \times 10^{-2}$. As we discuss all
Reed-Muller error correction to leading order only, we take the
leading order estimate
\begin{equation}
  \label{RMT}
  p_{Z,\text{c}}= \frac{1}{105} \approx 0.95 \times 10^{-2}.
\end{equation} 
For Error Model 1, from (\ref{eqZerr}) and (\ref{RMT}) we
obtain the threshold
\begin{equation}
  \label{RMT_M1}
  p_{1,\text{c}} = \frac{1}{70} \approx 1.4 \times 10^{-2},
  \;\;\;\;\;\;\mbox{(in $S$)}. 
\end{equation}
For error Model 2 we obtain
\begin{equation}
  \label{TRM_M2}
  \begin{array}{rclcl}
    p_{P,\text{c}}=p_{S,\text{c}}=p_{M,\text{c}}&=& \displaystyle{\frac{1}{350}
    \approx 0.29 \times 10^{-2}}, && \mbox{for }p_2=0, \vspace{1mm}\\ 
    p_{2,\text{c}}&=& \displaystyle{\frac{1}{532} \approx 0.19 \times
    10^{-2}}, && \mbox{for }p_P=p_S=p_M=0,\vspace{1mm}\\ 
    p_{i,\text{c}}&=& \displaystyle{\frac{1}{882} \approx 0.11 \times
    10^{-2}},  && 
    \mbox{for }p_P=p_S=p_M=p_2, 
  \end{array}
  \;\;\;\;\;\;\;\;\mbox{(in $S$)}.
\end{equation}
The topological error correction in $V$ and the Reed-Muller error
correction in $S$ run separately, and all that remains is to check 
which mechanism breaks down first. 
By comparison of Equation (\ref{Trpgm_M1}) with (\ref{RMT_M1}) and
Equation (\ref{Trpgm_M2}) with (\ref{TRM_M2}), we find for both error
models that the Reed-Muller error correction collapses first. It therefore
sets the overall threshold. In Error Model 1, the critical
error probability for local depolarizing error is $1.4 \times
10^{-2}$. In Error Model 2, for the case where preparation, gate,
storage and measurement errors each have equal strength, the error
threshold for the individual processes is $1.1 \times 10^{-3}$. 

\section{Overhead}
\label{O}

Denote by $N$ the number of qubits in the algorithm-specific resource state
$|\Psi_\text{algo}\rangle_Q$. $N$ is also the number of
non-Clifford one-qubit rotations in a quantum circuit realizing the
algorithm ``algo''. What is the number
$N_\text{ft}$ of cluster qubits required to perform the same
computation fault-tolerantly?

The qubit overhead factor for Reed-Muller error correction is
\begin{equation}
  O_\text{RM}= \left( \frac{\log N}{\log
  p_\text{c}/p}\right)^{\gamma}, 
\end{equation}
where $\gamma=\log_2 15 \approx 3.91$, and $p$ is the actual
and $p_\text{c}$ the critical Reed-Muller error rate. The set $S$
consists of $|S|=N O_\text{RM}$ qubits. 

We need to determine the
additional overhead due to topological error correction. We choose a
separation $r$ between strands of a defect loop, and a strand diameter
of $r/2$.
\begin{eqnarray}
  \parbox{5cm}{\epsfig{width=5cm, file=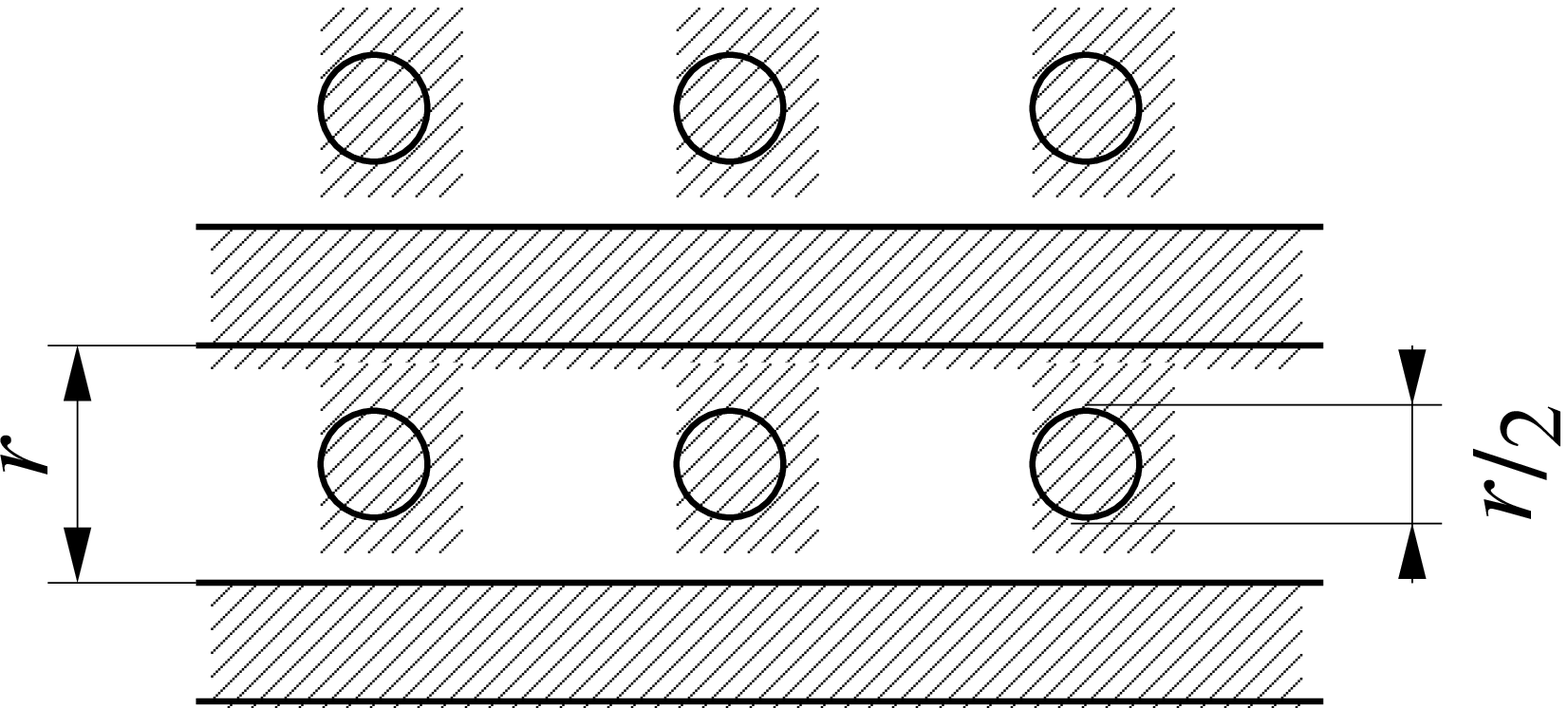}} \nonumber
\end{eqnarray}
The dimensions of the cluster thus are $\mbox{width}, \mbox{depth}
\leq 3/2\, r N O_\text{RM}$, $\mbox{height}=4r$. The most likely
errors then occur with a probability of $\exp(-\kappa(p) r)$,
(\ref{expsupp}), and there 
are less than $3r (N O_\text{RM})^2$  locations for them
(primal+dual). 
We require that
$3r (N O_\text{RM})^2 \exp(-\kappa(p) r) \approx 1$, such that
in the large $N$ limit
\begin{equation}
  r \approx \frac{2}{\kappa(p)} \ln N. 
\end{equation}     
Double-logarithmic corrections are omitted. The total number
of qubits is given by $N_\text{ft}=\mbox{width}\times
\mbox{depth}\times \mbox{height} \leq 9r^3(N O_\text{RM})^2$,
such that
\begin{equation}
  N_\text{ft} \sim  N^2 \left(\log N\right)^{3+2\gamma}.
\end{equation}
The overhead is polynomial. The values of the exponents in
the above expression may be reduced in more resourceful adaptations of
the presented scheme. 
 
\section{Discussion}
\label{Disc}

We have described a scheme for fault-tolerant cluster state universal quantum
computation which employs topological error
correction.  This is possible because of a link between cluster
states and surface codes. In addition to the topological method, we
make use of a Reed-Muller 
quantum code which ensures that non-Clifford operations can be performed
fault-tolerantly by local measurements. 

The error threshold is $1.4\%$ for an  
ad-hoc error model with local depolarizing error and $0.11\%$ for a more
detailed error model with
preparation-, gate-, storage- and measurement errors. We have not
tried to optimize for either threshold value or overhead here; the
foremost purpose of this 
paper is to explain the techniques. With regard to a high
threshold, the obvious bottleneck is the Reed-Muller code. The error threshold
imposed by this code on the cluster region $S$  is---depending on the
error model---a factor of 3 
to 5 times worse than the threshold obtained from the topological
error correction 
in $V$. To increase the threshold one may replace this code by
another CSS code with property (\ref{trans}) that has a higher error
threshold,
provided such a code exists. Alternatively, one may probe the
Reed-Muller code in error detection, as in magic state
distillation \cite{magic}. The error detection threshold is 
$14\%$, which indicates that there is some room for
improvement. 

Part of the investigations in this paper are numerical simulations,
and we would like to comment on their impact on the threshold
value.  Numerics are encapsulated only in the 
threshold estimate for topological error correction which is much
higher than the overall threshold. Our final threshold estimate stems
from the Reed-Muller code and is analytical.

\paragraph{Acknowledgments:}We would like to thank Sergey Bravyi,
Frank Verstraete, Alex McCauley, Hans Briegel and John Preskill for
discussions.   
JH is supported by ARDA under an Intelligence Community Postdoctoral
Fellowship.  KG is supported by DOE Grant No. DE-FG03-92-ER40701. RR
is supported  by MURI under Grant No. DAAD19-00-1-0374 
and by the National Science Foundation under contract number
PHY-0456720. Additional support was provided by the Austrian Academy of
Sciences. 

\vspace{5mm}

\appendix

\noindent
The appendices are relevant for Error Model 2 only.\vspace{-2mm}

\section{Connecting sub-clusters}
\label{SubClusters}

The cluster ${\cal{C}}$ consists of a set of sub-clusters
${\cal{C}}_k$, ${\cal{C}} = \bigcup_k {\cal{C}}_k$ which are prepared
in sequence. Two successive sub-clusters ${\cal{C}}_k$,
${\cal{C}}_{k+1}$ have an overlap, ${\cal{C}}_k \cap {\cal{C}}_{k+1}
=H_k \subset V \cup D$. $H_k$ is a set of locations for hand-over qubits.
Each graph edge (corresponding to a $\Lambda(Z)$-gate in cluster state
creation) can be unambiguously assigned to one sub-cluster. The set of
edges ending in one vertex either belongs to one or to two
sub-clusters. In the latter case, the vertex is the location for a
hand-over qubit. 

\begin{figure}
  \begin{center}
    \begin{tabular}{lcl}
      a) & b)\\
      \epsfig{width=2.3cm, file=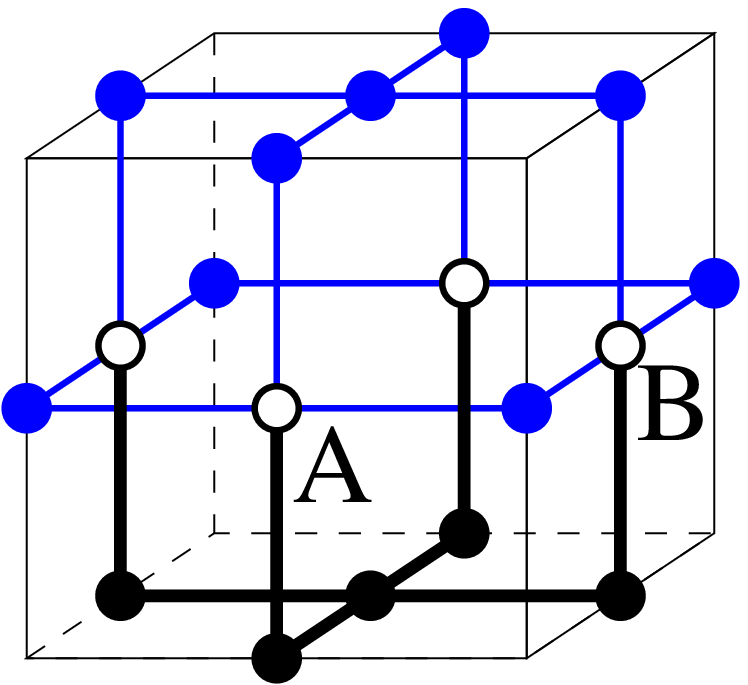} &&
      \epsfig{width=6.4cm, file=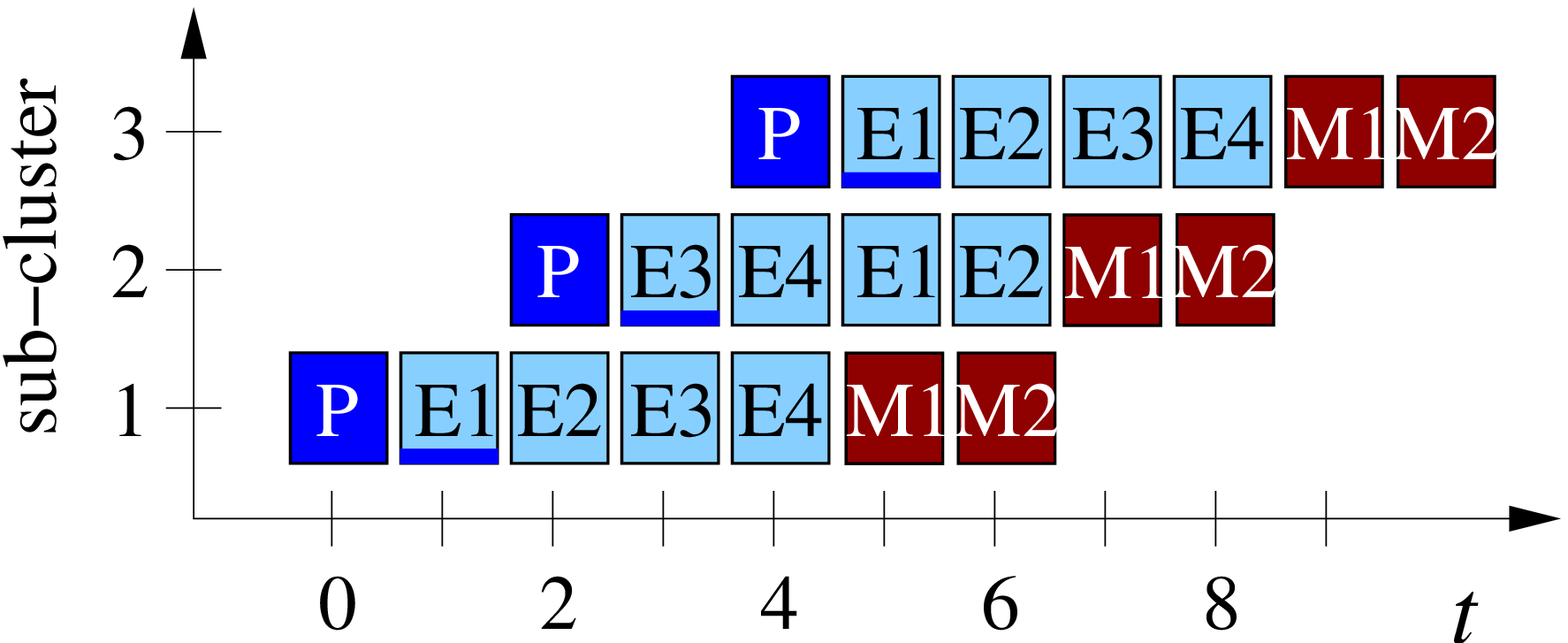}
    \end{tabular}
    \caption{\label{ssc}Connecting sub-clusters. a) Elementary cell of
    ${\cal{L}}$. The old sub-cluster is displayed in
    blue, the new in black;  ``$\circ$''; hand-over qubits,
    ``$\bullet$'': qubits within one
    sub-cluster. b) Temporal order of operations. ``P'':
     $|+\rangle$-preparation, ``E1-E4'': steps of parallel
    $\Lambda(Z)$-gates, M1, M2:  measurement.}
  \end{center}
\end{figure}

If $k$ is odd, the cluster state creation procedure on ${\cal{C}}_k$
is the sequence $1\rightarrow 2 \rightarrow 3 \rightarrow 4$. If $k$
is even, the sequence is   $3 \rightarrow 4 \rightarrow 1 \rightarrow
2$. As shown in Fig.~\ref{ssc}b, connecting the sub-clusters proceeds
smoothly.  Consider,
for example, the sub-cluster ${\cal{C}}_1$. With the exception of a
subset of the hand-over qubits to ${\cal{C}}_2$, the qubits in
${\cal{C}}_1$ are prepared at $t=0$, entangled in steps 1 to 4, and
measured at $t=5,6$. Specifically, the $V$- and $D$-qubits are
measured at time $t=5$, while the $S$-qubits are measured at time
$t=6$ with measurement
bases adapted according to previous measurement outcomes. The
sub-clusters ${\cal{C}}_k$ are chosen such that there is no temporal
order among the measurements on qubits within $S \cap {\cal{C}}_k$,
for all $k$. Then, the $S$-qubits wait for one time step in which a
storage error may occur.  

The hand-over qubits stay in the computation no longer than the other
$V$- and $D$-qubits. They form two subsets, $H_1 = A \cup B$; see 
Fig.~\ref{ssc}a. The qubits in $A$ are prepared at $t=0$, entangled in
time steps 1 to 4 and measured in at $t=5$. They cause no change in the
error model. 

The qubits in $B$ are not acted upon by a gate until time
step 2 (since the potential interaction partner of step 1 isn't there yet),
so they are prepared at time $t=1$. The final interaction involving
the $B$-qubits is in step 5, and they are measured in step
6. Between preparation and measurement, the $B$-qubits are in the
computation for four time steps in each of which they are acted upon
by a gate. No additional storage error occurs. There is one 
modification due to the $B$-qubits. The temporal order of
$\Lambda(Z)$-gates involving the qubits $b \in B$ is changed. As a
result, the correlated errors on the edge qubits of the faces
$\{c_2\}=b$ are not among pairs of opposite edge qubits but among
pairs of neighboring
edge qubits. So, the error rates $q_1$ and $q_2$ in (\ref{RPGMmap2})
are unchanged, but $q_2$ characterizes a slightly different process. 

We expect this to be a minor effect. The overall threshold
is still set by the threshold for Reed-Muller error-correction, which
is some five times smaller than the simulated threshold for
topological error correction.

\section{Effective error channel on the $S$-qubits}
\label{E2}

\begin{figure}
  \begin{center}
    \epsfig{width=7.5cm, file=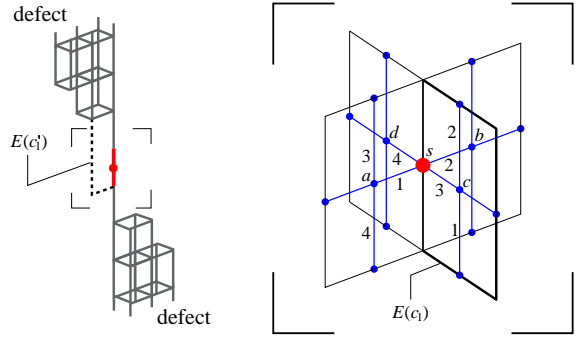}
    \caption{\label{SEC}Lattice around an $S$-qubit. Left: Location of
    the $S$-qubit (red) and the defect (gray). $E(c_1^\prime)$ is an error
    cycle of length 3 and weight 3 for Error Models 1 and 2.  Right: Detail.
    $E(c_1)$ is a non-trivial error cycle of length 3 which has weight
    2 in error Model 2. Red: $S$-qubit, black:
    lattice ${\cal{L}}$ for topological error correction. 
    The cluster edges (blue) correspond to $\Lambda(Z)$-gates whose
    temporal order is as indicated.}
  \end{center}
\end{figure}

The effective error on an $S$-qubit stems from the
$S$-qubit itself and its immediate surrounding shown in Fig.~\ref{SEC}
and from the two
edge-qubits in the one-dimensional section of the defect, which are
not protected by any syndrome. Of the latter each contributes an error
\begin{equation}
  \label{Defe}
  E_\text{defect} =
  \left(\frac{4}{5}p_2+\frac{2}{3}p_S+\frac{2}{3}p_M\right) [X_s].
\end{equation}
The preparation error does not contribute, because the corresponding
$Z$-error on the defect qubit is absorbed in the $Z$-measurement. 

The effective error of the center qubit $s \in S$ stems from
operations that act on $s$ directly, from $X$- or $Y$-errors propagated to $s$
by the $\Lambda(Z)$-gates  and from short nontrivial
cycles. Specifically, there are four non-trivial cycles of length
3. One of them is denoted as $E(c_1)$ in Fig.~\ref{SEC}. Because of the
correlations in the forward-propagated errors these cycles have weight
2 and cause inconclusive syndrome at lowest order in the error
probability. There are further error cycles of length 3,
such as $E(c_1^\prime)$ in Fig~\ref{SEC}. But they have weight 3 even for
Error Model 2 and do not contribute to the lowest order error channel. 
We perform a count including all error sources
in the cluster region displayed in Fig.~\ref{SEC}, right. There is
one convention that enters into the count. Namely, the error
$Z_aZ_bZ_cZ_d$ (see  Fig.~\ref{SEC})  is a non-trivial error
cycle such that the errors $Z_aZ_b$ and $Z_cZ_d$ have the same weight and
the same syndrome but different effect on the computation, $X_s$
vs. $I_s$. When the 
corresponding syndrome occurs, we assert that a logical $X$-error
occurred and correct for it. We obtain
\begin{equation}
  \label{Cene}
  E_\text{central}=\left(\frac{2}{5}p_2 + \frac{1}{3}p_S +\frac{1}{3}p_M
  \right)\left([X_s]+[Y_s]\right)+
  \left(\frac{2}{3}p_P+\frac{1}{3}p_S+\frac{1}{3}p_M+\frac{58}{15}p_2\right)
  [Z_s]. 
\end{equation}
The sources (\ref{Defe}) and (\ref{Cene}) combined,
$E_\text{central}+2\times E_\text{defect}$, lead to the local error
channel (\ref{effSEC}). 
 
\section{Conversion of $X$- and $Y$-errors on $S$-qubits}
\label{E3}
Here we show that an $X$- or $Y$-error on an individual $S$-qubit $s$ with
probability $p$ is equivalent to a $Z$-error on that qubit with
probability $p/2$, $p [X_s] \cong p[Y_S] \cong p/2\,
[Z_s]$. The qubit $s$ may be measured in the eigenbasis of
$\frac{X+Y}{\sqrt{2}}$ or of $\frac{X-Y}{\sqrt{2}}$. W.l.o.g. assume the
qubit $s$ is measured in the eigenbasis of 
$\frac{X+Y}{\sqrt{2}}$. Then, $X_s=\frac{X_s+Y_s}{\sqrt{2}}
\frac{I_s - iZ_s}{\sqrt{2}} \cong \frac{I_s - i Z_s}{\sqrt{2}} $. The
$X$-error is equivalent to a coherent $Z$-error and we need to check
whether the coherences matter. More generally, for the described
scenario with the subsequent measurement a probabilistic local error
channel $(1-p_X-p_Y-p_Z)[I_s]+p_X[X_s]+p_Y[Y_s]+p_Z[Z_s]$ is
equivalent to a channel with coherent errors
\begin{equation}
  \label{CohE}
  \rho \longrightarrow (1-q) \rho + q Z_s \rho Z_s + i \tilde{q}
  \left(\rho Z_s - Z_s \rho \right),
\end{equation}
with
\begin{equation}
  \label{hp}
  q = p_Z+ \frac{p_X+p_Y}{2},\;\;\; \tilde{q} = \frac{p_X - p_Y}{2}.
\end{equation}
Now assume that all $S$-qubits are affected individually by the
error channel (\ref{CohE}). The Reed-Muller error correction at
successive levels maps these channels to channels of the same form, one coding
level higher up. The parameters $q_l$, $\tilde{q_l}$ at coding level
$l$ obey recursion relations which, up to fourth order, read 
\begin{equation}
  \label{MRC}
  \begin{array}{lcl}
    q_{l+1} &=& \displaystyle{105\, q_l^2\, (1-q_l)^{13} + 35\, q_l^3\,
    (1-q_l)^{12}+ 1260\, 
    q_l^4\,(1-q_l)^{11}+ 630\, \tilde{q}_l^4\,(1-q_l)^{11}},\vspace{1mm}\\
    \tilde{q}_{l+1} &=& \displaystyle{70 \,\tilde{q}_l^3
    \,(1-q_l)^{12}-1680\, \tilde{q}_l^3 q_l \, (1-q_l)^{11}}.
  \end{array}
\end{equation} 
If we compare
(\ref{MRC}) to the recursion relation of $q$ for probabilistic
$Z$-error, a deviation first shows up at fourth order. The discussion
of Reed-Muller error correction in this paper is confined to leading
order.
The leading order result for the threshold, $q_\text{c}=1/105$,
is not affected by the coherences in the error (\ref{CohE}).
Probabilistic $X$-and $Y$-errors influence the threshold by
contributing half their weight to the probability of an effective
$Z$-error; see (\ref{hp}).  

\end{document}